\documentclass[journal=nalefd,manuscript=letter]{achemso}

\usepackage{achemso}
\usepackage{graphicx} 
\usepackage{dcolumn}
\usepackage{bm}
\usepackage{booktabs}
\usepackage{textcomp}
\usepackage{makecell}
\usepackage{hhline}
\usepackage[usestackEOL]{stackengine}
\usepackage[version=3]{mhchem}
\usepackage{amsmath}
\usepackage{amsfonts}
\usepackage{braket}

\usepackage{multirow}

\usepackage[caption=false]{subfig}
\usepackage{tikz}

\title{Low-energy nine-layer rhombohedral stacking of transition metal dichalcogenides}
\author{Rijan Karkee} 
\affiliation{Department of Physics, University of California, Merced, CA 95343}

\author{David A. Strubbe} 
\email{dstrubbe@ucmerced.edu}
\affiliation{Department of Physics, University of California, Merced, CA 95343}

\date{\today}

\begin{document}

\section{Abstract}\label{Abstract}
Transition-metal dichalcogenides (TMDs) show unique physical, optical, and electronic properties. The known phases of TMDs are 2H and 3R in bulk form, 1T and associated reconstructions, and 1H in monolayer form. This paper reports a hypothetical phase, 9R, that may exist in TMDs (Mo, W)(S, Se, Te)$_2$, meeting both dynamical stability and elastic stability criteria. 9R phase has the same space group as 3R, $i.e.$ rhombohedral $R3m$ without inversion symmetry, and has 9 layers in a conventional unit cell. We find that 9R has an energy within 1 meV per formula unit of 3R and can be energetically favored by a particular strain condition. We further calculate the electronic, elastic, piezoelectric, Raman, and second-harmonic generation signatures of 9R TMDs and compare them with the corresponding 2H and 3R phases. 9R has similar properties to 3R but shows distinctive Raman peaks in the low-frequency regime, improved piezoelectric properties, and unique band splitting arising from layer coupling at the conduction band minimum. These distinct properties make 9R an attractive candidate for applications in piezotronics and valleytronics.

\section{Introduction}

Transition-metal dichalcogenides (TMDs) have emerged as a class of attractive two-dimensional materials and have attracted significant attention in the fields of condensed matter physics and nanotechnology. TMDs exhibit a layer structure consisting of transition metal atoms, usually in groups 4 to 7, sandwiched between Chalcogen atoms such as Sulfur, Selenium, or Tellurium. The different atomic arrangements and reduced dimensions of TMDs produce a wealth of unique electronic \cite{appl_electronic_tmd}, optical \cite{appl_optic_tmd} and mechanical properties \cite{appl_mechanical_tmd}, attracting considerable interest in their fundamental research and potential technological applications. Due to their adjustable band gaps \cite{gap_variation}, strong spin-orbit couplings \cite{spin_orb_tmd}, and high carrier mobility \cite{mobility}, TMDs have proved promising capabilities in the fields of optical, electronics, catalysis, and energy storage. In recent years, significant progress has been made in synthesizing and characterizing TMD monolayers and heterostructures, revealing new phenomena and enabling the engineering of multifunctional devices \cite{FET_tmd,diodes_tmd}. Research in this field is ongoing, and the characterization of complex TMD behaviors and their unique characteristics has a profound impact on the development of next-generation electronic and photonic technologies.

Furthermore, to add the prospect, TMDs come in different phases $i.e.$ monolayer and bulk offering further interesting properties. For example, in MoS$_2$, monolayer 1T is metallic while monolayer 1H is semiconducting with a direct gap \cite{Karkee2024}; and 2H and 3R are bulk phases that differ by stacking (sliding followed by rotation) but are similar in energetics and electronic bandstructure with indirect band-gap from theoretical calculation \cite{Karkee2020}, yet 2H is centrosymmetric while 3R is noncentrosymmetric offering different applications such as piezoelectricity and second harmonic generation \cite{3R_review}. Given the importance of other phases might bring new insight into the application, it is important to look for other polymorphs. In this paper, we report a nine-layer phase belonging to R3m space group and because of this, we call 9R, as per Ramsdell notation, hereafter as a potential new phase of TMDs. The 9R phase is a known phase for other compounds like SiC \cite{9R_SiC} but not reported for TMDs. A literature investigated hypothetical polytypes of MoS$_2$ based on a random searching method using density functional theory reported several stable and unstable polymorphs of MoS$_2$ but did not report the nine-layer phase. \cite{hypothetical_phase_tmd} In that paper, an unstable 6R phase is reported. 

In this paper, we propose the nine-layer phase in 6 TMDs -- MoS$_2$, MoSe$_2$, MoTe$_2$, WS$_2$, WSe$_2$, and WTe$_2$ -- meeting the criteria of dynamical stability and elastic stability. We then discuss the unique signatures in Raman, SHG and diffraction patterns for the new phase and explore some properties, like piezoelectricity and spin-orbit splitting of bands. Finally, we suggest possible experiments that would allow us to synthesize the new phase via controlling strain. 

\section{Methodology}
\label{Methods_4}
Our calculations use plane-wave density functional theory (DFT)  implemented in the code Quantum ESPRESSO, version 7.1  \cite{QE-2009,QE_2}. We used the Perdew-Burke-Ernzerhof (PBE) generalized gradient approximation  \cite{PBE}  for structural analysis and electronic bandstructure, and local density approximation (LDA)   functional  \cite{lda_perdew} for Raman, elastic and peizoelectric properties. 
With PBE, we used the semi-empirical  Grimme-D2 (GD2)  \cite{Grimme} Van der Waals correction to the total energy, which gives lattice  parameters  and other properties considerably closer to experimental results \cite{Mos2_exp_fit,enrique}. Calculation with LDA has also been shown to give accurate lattice parameters  \cite{enrique}. We used Optimized Norm-Conserving Vanderbilt pseudopotentials  \cite{ONCV}  and fully relativistic pseudopotential set for spin-orbit calculation  from Pseudodojo set \cite{web_pseudojo}. Kinetic energy cutoffs of 952 eV (70 Ry) for PBE and 1224 eV (90 Ry) for LDA were used.  Half-shifted $k$-point grids of $6\times6\times2$  were chosen to converge the total energies within 0.001 eV/atom for 2H, 3R and 9R. Atomic coordinates were relaxed using a force threshold of $10^{-4}$ Ry/bohr and the stresses were relaxed below $0.1$ kbar. For calculating the piezoelectricity,  we used the Berry-phase method \cite{RevModPhys} for calculation of polarization, where we used 30 k-points (\texttt{nppstr}=30) in the direction of k-point strings (\texttt{gdir}).

\section{Results}

\subsection{Structure, energetics, and stability}
\begin{figure}[H]
\includegraphics[scale=1.87]{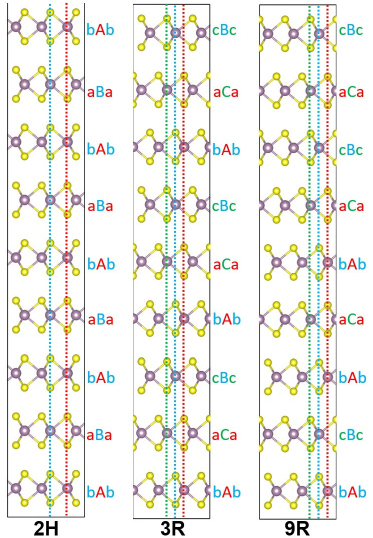}
\caption{Structure comparison in 2H, 3R and 9R phase. The dotted red, blue and green lines corresponds to  A, B  and C alphabets and transition metal layer is represented by capital letters and sandwiching chalcogens are represented by small letters.}
\label{fig:9r_structure}
\end{figure}

The structure of the 9R phase has the same space group of 3R $i.e. R3m$ containing nine layers with 27 atoms in a conventional unit cell. The stacking sequence of the structure is bAb cBc bAb aCa bAb aCa cBc aCa cBc, where small letter represents the transition metal and capital letter represents the chalcogens as shown in Figure \ref{fig:9r_structure}.  The details of the lattice parameters comparison are shown in Table \ref{tab:9r_lattice_parameters_comp}. The conventional unit cell of 9R has $a=b \neq c$ and $\alpha=\beta= 90$\textdegree and $\gamma=120$\textdegree. Also, note that there exists a primitive unit cell for 9R phase with 9 atoms per unit cell. In the primitive cell, $a=b=c$ and $\alpha =\beta =\gamma \neq 90$\textdegree.

\begin{table}[H]
    \centering
    \begin{tabular}{|c|r|r|r|r|r|r|r|r|r|}
    \hline
      \multicolumn{1}{|c|}{\multirow{2}[0]{*}{\textbf{{{TMDs} }}}} 
 & \multicolumn{3}{c|}{\textbf{$a=b$}} & \multicolumn{3}{c|}{\textbf{$c$}} & \multicolumn{3}{c|}{\textbf{Layer spacing}}\\
 
  \cline{2-10}  
  
\multicolumn{1}{|c|}{} & \multicolumn{1}{c|}{{2H}}  & \multicolumn{1}{c|}{{ 3R}} & \multicolumn{1}{c|}{{  9R}}  & \multicolumn{1}{c|}{{ 2H}} & \multicolumn{1}{c|}{{ 3R}} & \multicolumn{1}{c|}{{  9R}} & \multicolumn{1}{c|}{{ 2H}} & \multicolumn{1}{c|}{{ 3R}} & \multicolumn{1}{c|}{{  9R}} \\
 \hline \hline
 MoS$_2$ & 3.18 & 3.19 & 3.19   & 12.36 & 18.47 &55.42 & 6.18 & 6.16 & 6.16  \\
\hline
MoSe$_2$ & 3.24  & 3.25 & 3.25& 12.18   & 18.06  &54.20 & 6.09 & 6.02 & 6.02  \\
\hline
MoTe$_2$ & 3.52  & 3.53 & 3.53   & 13.96 &20.95 & 62.87 & 6.98 & 6.97 & 6.97\\
\hline
WS$_2$ & 3.18  & 3.18 & 3.18   & 12.14 & 18.14 &54.43& 6.08 & 6.04 & 6.05\\
\hline
WSe$_2$  & 3.24  & 3.26 & 3.26   & 12.04 &17.86 &53.61& 6.02 & 5.95 & 5.95  \\
\hline
WTe$_2$  & 3.56  & 3.56 & 3.56  & 13.78 & 20.68 &62.07& 6.89 & 6.89 & 6.90 \\
\hline

    \end{tabular}
    \caption{Lattice parameters comparison in 2H, 3R and 9R phase in a conventional unit cell for 3R and 9R using PBE+GD2. All units are in \AA\rm. }
    \label{tab:9r_lattice_parameters_comp}
\end{table}

Table \ref{tab:9r_energy_diff} shows the energy difference between the 2H, 3R, and 9R phases in 6 TMDs with PBE+GD2 and LDA. We see that the energy difference between 3R and 9R phase is very small, less than 1 meV per formula unit for PBE+GD2. This energy difference is generally less than the difference between 2H and 3R, indicating the thermodynamic accessibility of 9R.

\begin{table}[H]
\centering
\caption{Energy difference between 2H, 3R and 9R phases (meV per formula unit). Here the energy of 3R is set to zero, so more the negative value, the more the phase is favorable. 2H phase is most favorable and 9R is least favorable phase, except in LDA calculation in MoTe$_2$, where 9R has lower energy than 3R. }
\begin{tabular}{|c|r|r|r|r|}

\cline{1-5}
  \multicolumn{1}{|c|}{\multirow{2}[0]{*}{\textbf{{{Structure} }}}} 
 & \multicolumn{2}{c|}{\textbf{PBE+GD2}} & \multicolumn{2}{c|}{\textbf{LDA}} \\
 
  \cline{2-5}  
  
\multicolumn{1}{|c|}{} & \multicolumn{1}{c|}{{$\Delta E_{\rm 2H - 3R} $}}  & \multicolumn{1}{c|}{{ $\Delta E_{\rm 9R - 3R} $}} & \multicolumn{1}{c|}{{  $\Delta E_{\rm 2H - 3R} $}}  & \multicolumn{1}{c|}{{ $\Delta E_{\rm 9R - 3R} $}}   \\
 \hline \hline

MoS$_2$ & -0.11  & 0.11 & -0.97   & 0.31  \\
\hline
MoSe$_2$ & -8.61  & 0.03 & -4.72   & 0.34  \\
\hline
MoTe$_2$ & -12.4  & 0.45 & -17.1   & -1.11  \\
\hline
WS$_2$ & -0.83  & 0.21 & -4.70   & 0.23  \\
\hline
WSe$_2$  & -5.57  & 0.33 & -8.62   & 0.24  \\
\hline
WTe$_2$  & -15.6  & 0.60 & -21.3   & 0.47  \\
\hline

\end{tabular}
\label{tab:9r_energy_diff}
\end{table}

\subsubsection{Dynamical stability}
\begin{figure}[H]
			\centering{
			\begin{tikzpicture}
			\node [anchor=north west] (imgA) at (-0.15\linewidth,.58\linewidth){\includegraphics[width=0.335\linewidth]{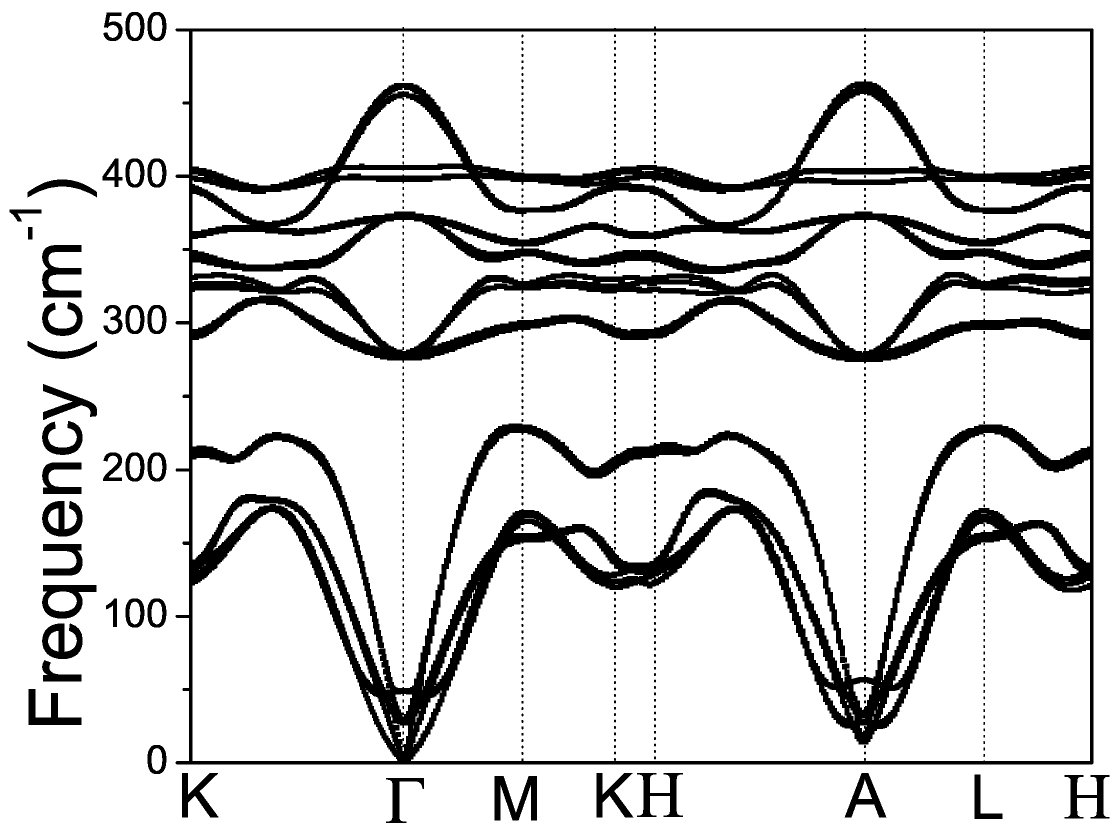}};
            \node [anchor=north west] (imgB) at (0.186\linewidth,.58\linewidth){\includegraphics[width=0.33\linewidth]{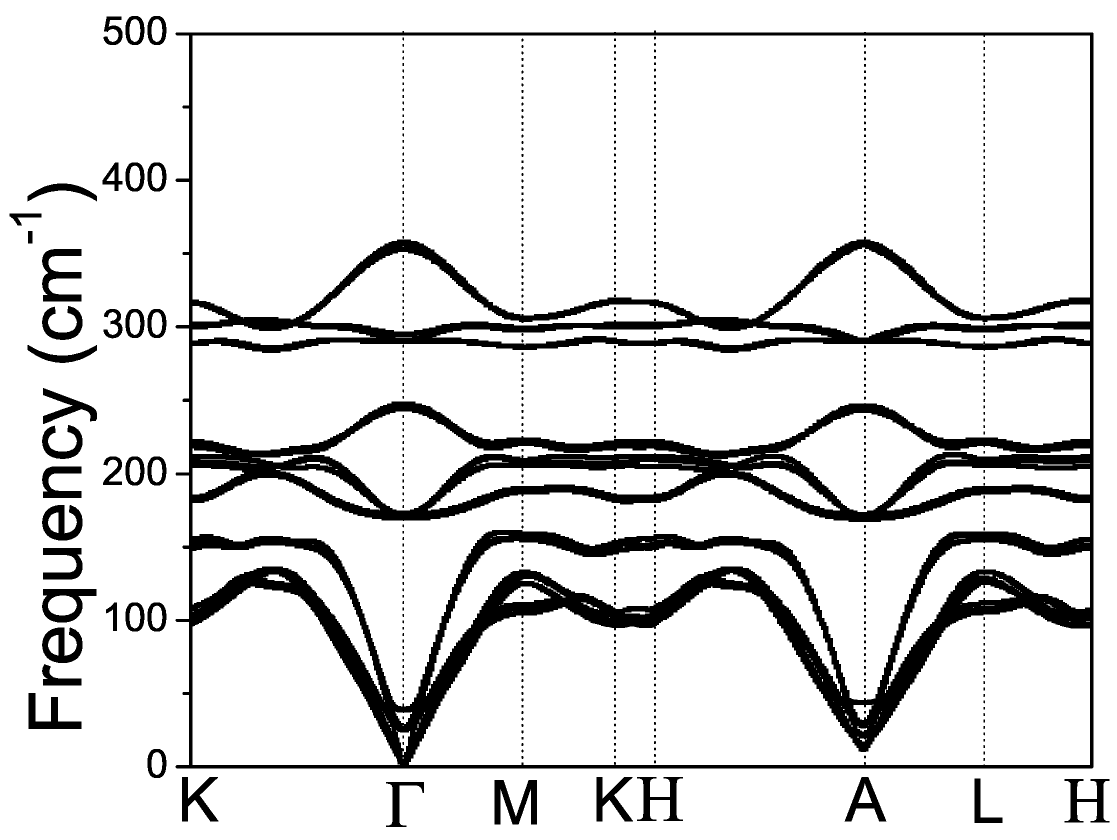}};
          \node [anchor=north west] (imgC) at (0.52\linewidth,.58\linewidth){\includegraphics[width=0.33\linewidth]{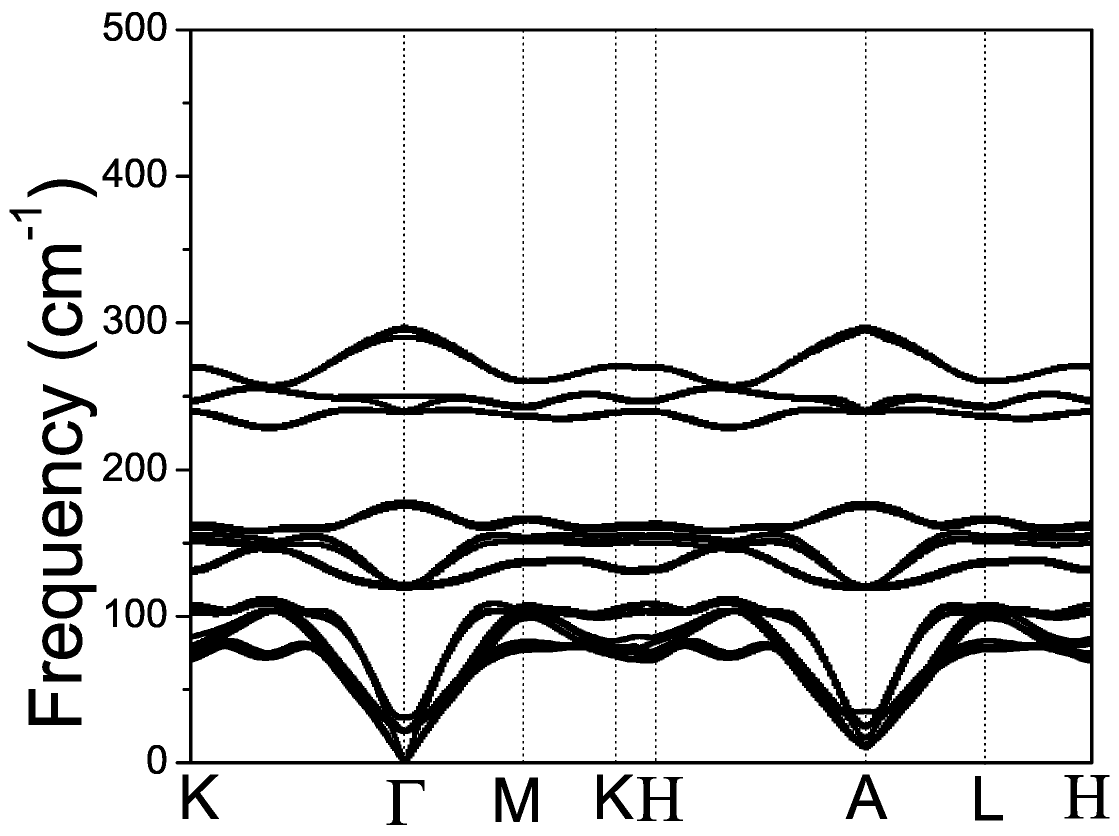}};
          
          	\node [anchor=north west] (imgD) at (-0.15\linewidth,.265\linewidth){\includegraphics[width=0.333\linewidth]{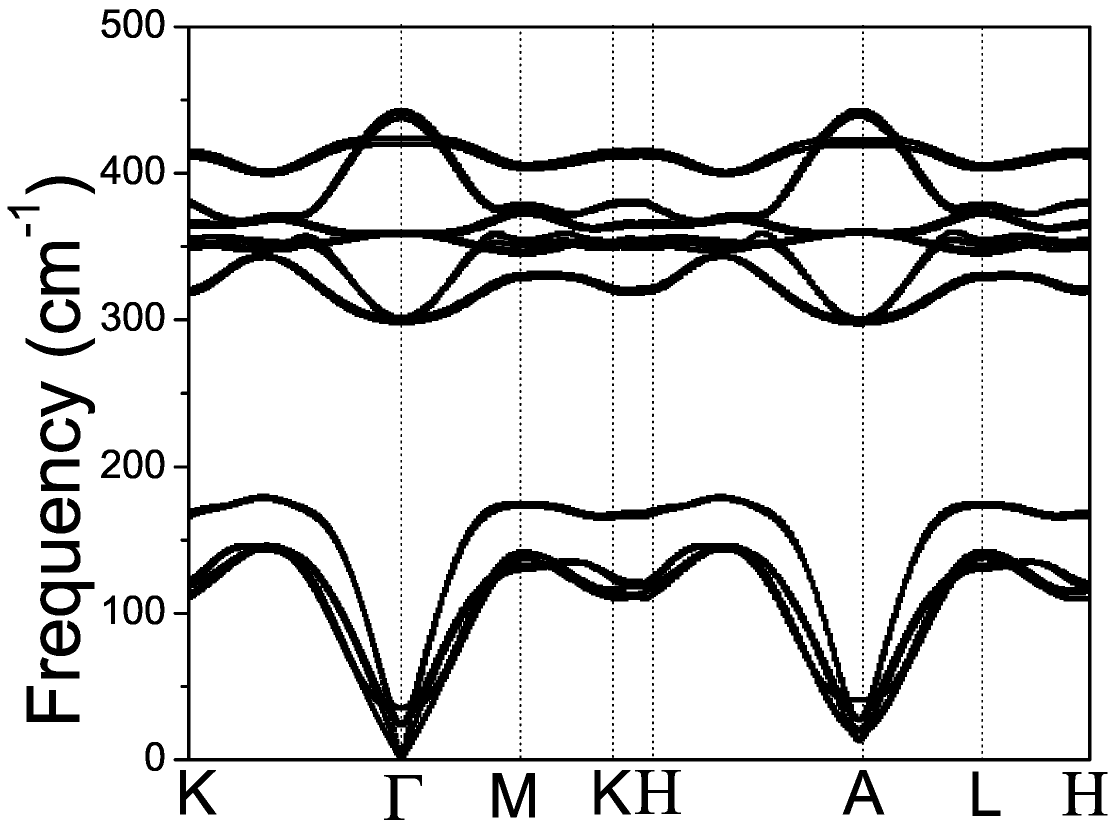}};
            \node [anchor=north west] (imgE) at (0.186\linewidth,.26\linewidth){\includegraphics[width=0.33\linewidth]{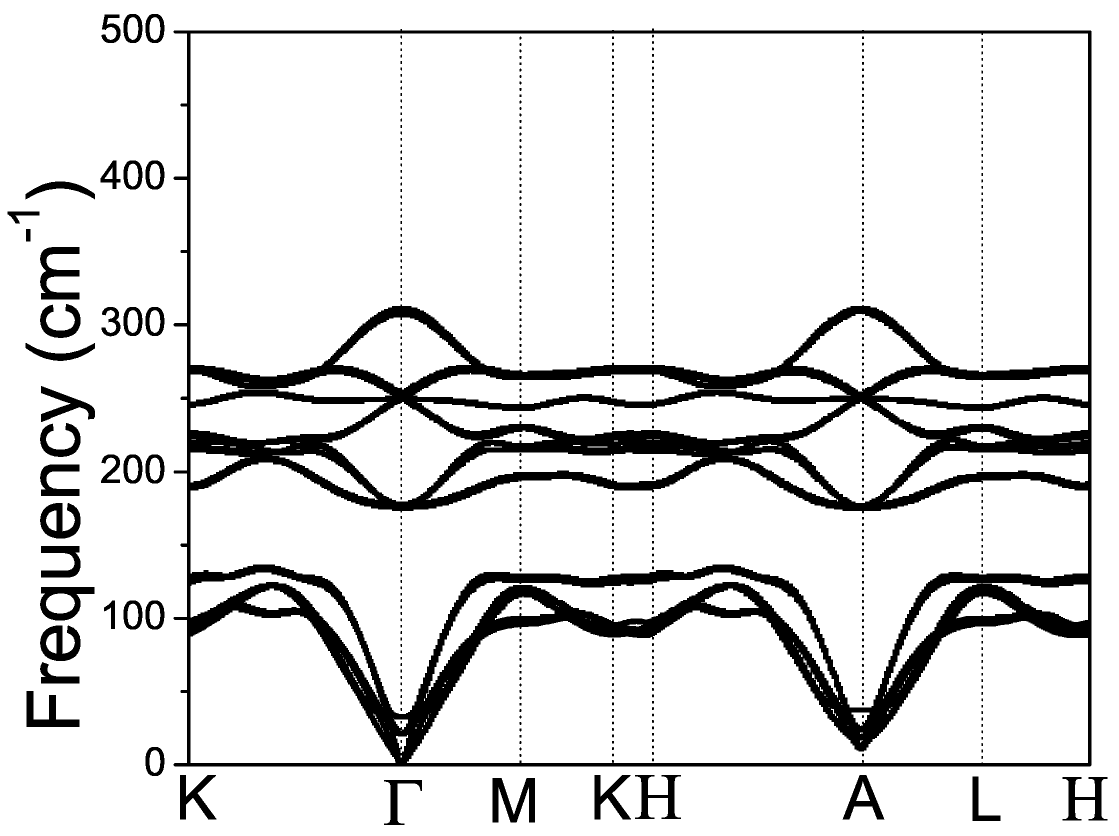}};
          \node [anchor=north west] (imgF) at (0.52\linewidth,.26\linewidth){\includegraphics[width=0.33\linewidth]{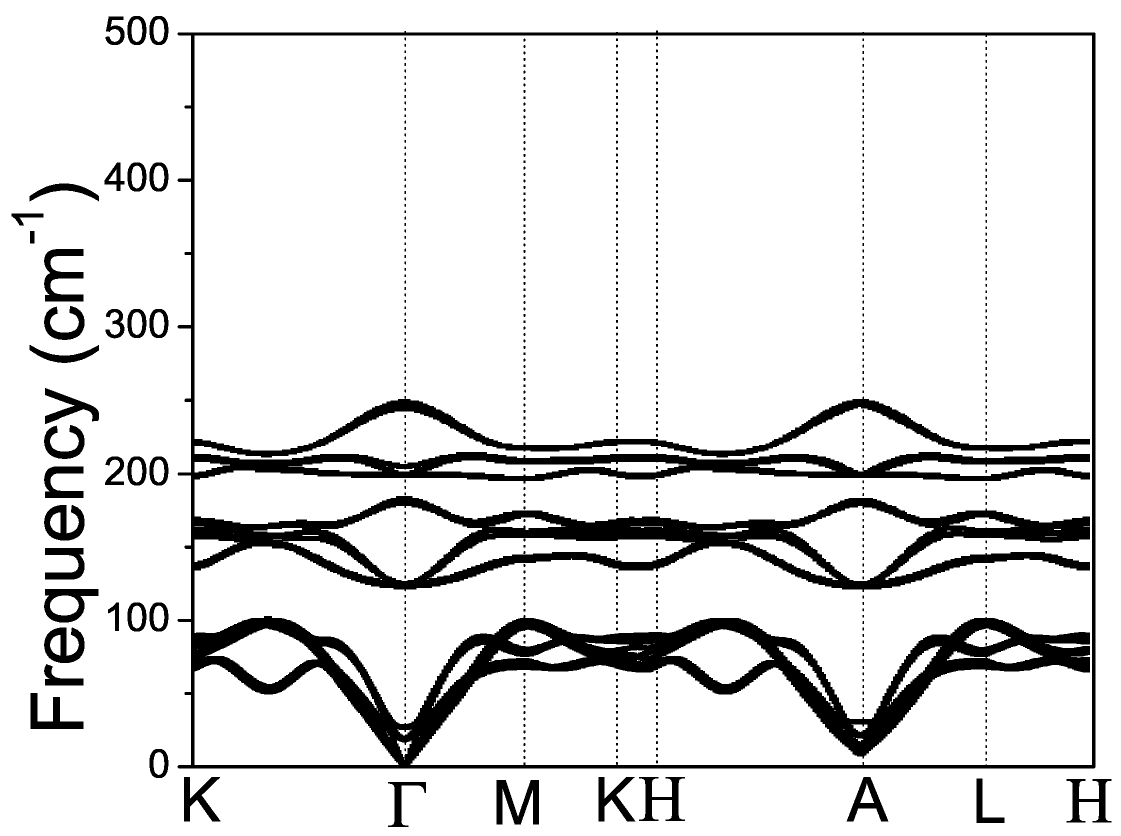}};
          
            \draw [anchor=north west] (-0.15\linewidth, .61\linewidth) node {(a) {\fontfamily{Arial}\selectfont \textbf{MoS$_2$}}};
            \draw [anchor=north west] (0.19\linewidth, .61\linewidth) node {(b) {\fontfamily{Arial}\selectfont \textbf{MoSe$_2$}}};
           \draw [anchor=north west] (0.52\linewidth, .61\linewidth) node {(c) {\fontfamily{Arial}\selectfont \textbf{MoTe$_2$}}};
            \draw [anchor=north west] (-0.15\linewidth, .30\linewidth) node {(d) {\fontfamily{Arial}\selectfont \textbf{WS$_2$}}};
            \draw [anchor=north west] (0.19\linewidth, .30\linewidth) node {(e) {\fontfamily{Arial}\selectfont \textbf{WSe$_2$}}};
           \draw [anchor=north west] (0.52\linewidth, .30\linewidth) node {(f) {\fontfamily{Arial}\selectfont \textbf{WTe$_2$}}};          
            \end{tikzpicture}
			}
			
			\caption{Calculated phonon bandstructure of 9R TMDs both in the in-plane and out-plane of the Brillouin zone.}
		\label{fig:9R_phonon}
			
		\end{figure}
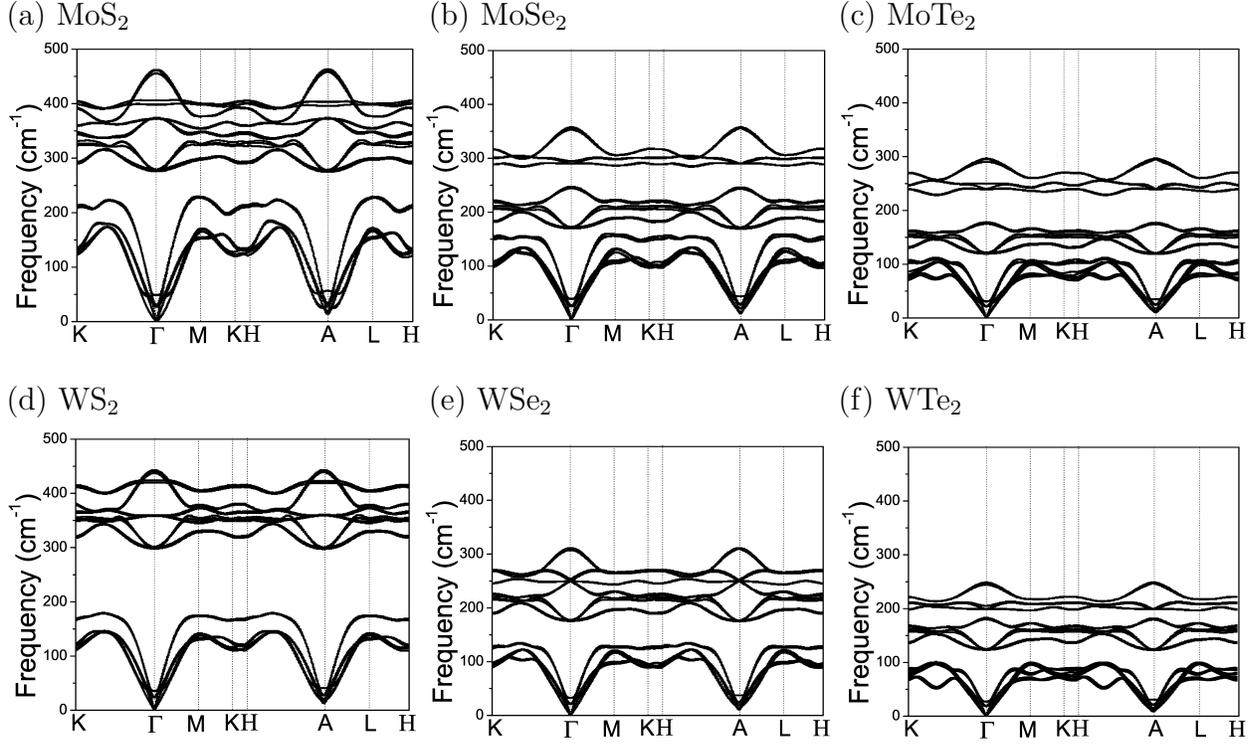

We tested the dynamical stability of the material by computing the phonon dispersion and found no imaginary frequencies in the dispersion for all TMDs. The presence of non-negative frequency modes corresponds to the structure being stable and at local minimum in the potential energy surface. So, under small perturbation, the structure can be retained after relaxation. Figure \ref{fig:9R_phonon} shows the phonon bandstructure of 9R TMDs in the in-plane and out-plane of the hexagonal Brillouin zone. For comparison, we also calculated the phonon bandstructure of 3R TMDs and results are similar and is shown in Supplementary Information (Fig \ref{fig:3R_phonon}).

\subsubsection{Elastic stability}
The crystal structure is stable in the presence of external loads within the harmonic approximation limit if and only if both the dynamical stability and elastic stability criteria are satisfied. In the previous section, we discussed the stability based on positive phonon frequencies across the Brillouin zone. The necessary and sufficient elastic stability criteria for the rhombohedral class of crystals are\cite{elastic_criteria}:
\begin{equation}
\label{EQN_elastic_criteria}
\begin{array}{l}
    C_{11} > \lvert C_{12} \rvert, \space  C_{44} > 0, \\
    C_{13}^2 < \frac{1}{2} C_{33} (C_{11}+C_{12}), \\
     C_{14}^2 < \frac{1}{2} C_{44} (C_{11}-C_{12})

\end{array}
\end{equation}

\begin{table}[htbp]

\centering
\caption{Elastic coefficients (GPa).}
\label{Tab_Elastic_constants}
\begin{tabular}{|cc|r|r|r|r|r|r|}
\hline
\multicolumn{2}{|c|}{Phase} & $C_{11}$ & $C_{12}$ &  $C_{13}$ & $C_{14}$ & $C_{33}$ & $C_{44}$ \\
\hline
\hline
 & 2H & 240 & 57 & 10 & 0 & 53 & 17 \\
 MoS$_2$ & 3R & 242 & 60 & 15 & 4 & 45 & 19 \\
  & 9R & 242 & 60 & 16 & 1 & 46 & 18 \\
\hline
\hline
 & 2H & 194 & 44 & 13 &  0& 52 & 17 \\
 MoSe$_2$ & 3R & 194 & 46 & 17 & 4 & 45 & 17 \\
  & 9R & 194 & 46 & 18 & 1 & 46 & 17 \\
\hline
\hline
 & 2H & 138 & 32 & 14 & 0 & 52 & 23 \\
 MoTe$_2$ & 3R & 138 & 35 & 18 & 3 & 42 & 18 \\
  & 9R & 137 & 34 & 18 & 1 & 42 & 17 \\
\hline
\hline
 & 2H & 261 & 55 & 10 & 0 & 52 & 22 \\
 WS$_2$ & 3R & 262 & 57 & 14 & 4 & 44 & 21 \\
  & 9R & 262 & 57 & 15 & 1 & 45 & 16 \\
\hline
\hline
 & 2H & 210 & 40 & 12 & 0 & 52 & 25 \\
 WSe$_2$ & 3R & 209 & 42 & 16 & 4 & 44 & 17 \\
  & 9R & 208 & 40 & 16 & 1 & 44& 15 \\
\hline
\hline
 & 2H & 147 & 25 & 13 & 0 & 52 & 23 \\
 WTe$_2$ & 3R & 146 & 28 & 17 & 3 & 42 & 17 \\
  & 9R & 145 & 27 & 16 & 1 & 42 & 16 \\
\hline

\end{tabular}
\end{table}

Based on Table \ref{Tab_Elastic_constants}, we can see that all elastic criteria as shown in Equation \ref{EQN_elastic_criteria} are satisfied for 9R. Many elastic constants of 2H and 3R in those TMDs are not reported. Some of them we found agree well with experimental or theoretical works. For example, 2H-MoS$_2$ agrees well with experiment \cite{FELDMAN19761141} and DFT work \cite{YuanChengZhangChenCai}. Similarly, the elastic tensor of 3R-MoS$_2$ agrees well with the theoretical work \cite{3R_review}. Furthermore, the elastic tensor of 2H-phase of MoSe$_2$ agrees well with the experiment \cite{MoSe2_elastic_exp}.

\subsection{Raman characterization of TMDs}

\begin{figure}[H]
    \centering
    \includegraphics[scale=0.5]{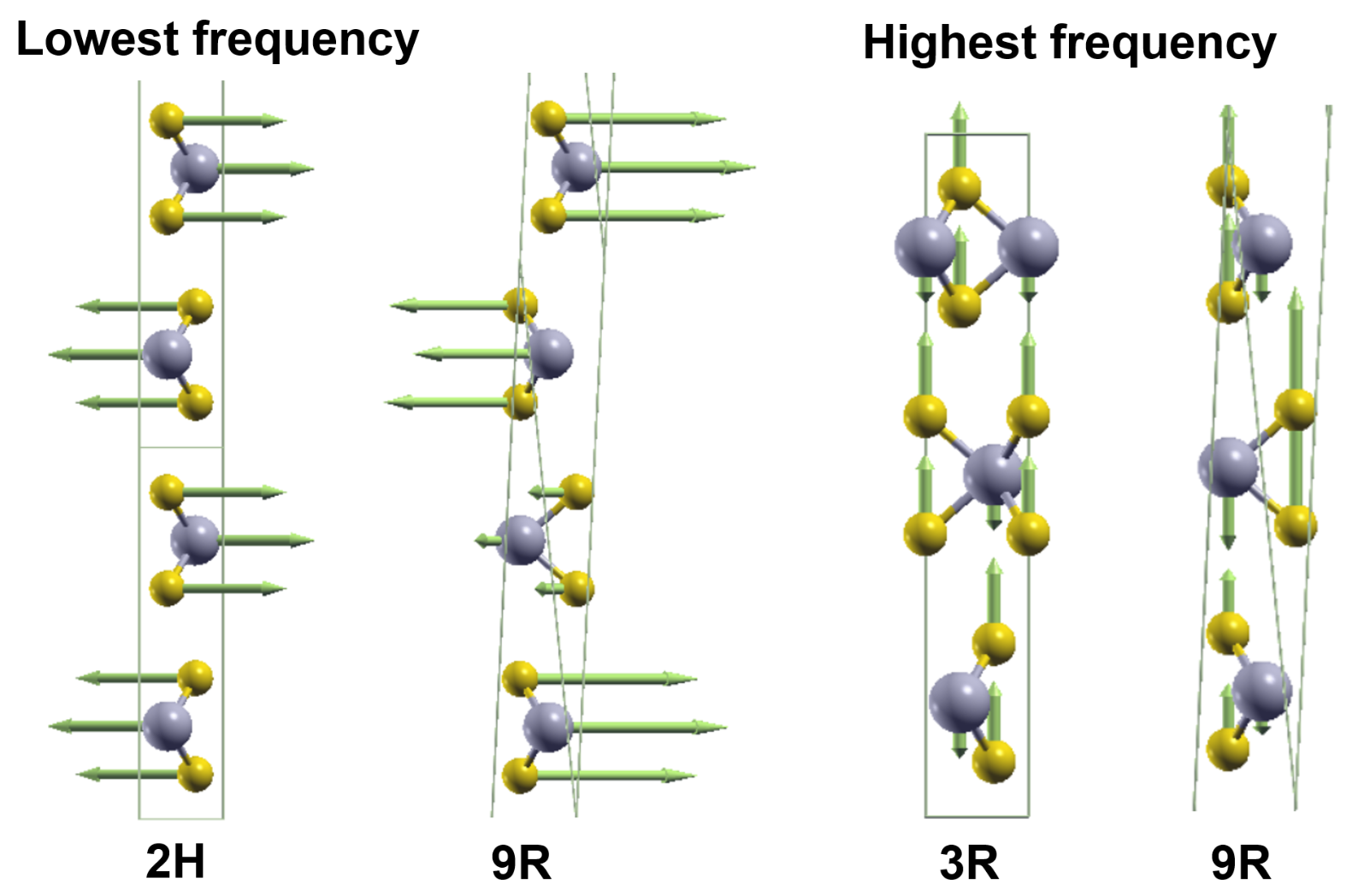}
    \caption{Modes of vibration in low and high Raman active modes. Here, transition metal is shown in grey and chalcogens are shown in yellow. The boxes represent primitive (2H and 9R) and a conventional cell (3R). Farthest Raman modes varies depending on material but this mode corresponds to the rightmost Raman active mode in Fig. \ref{fig:Raman_9R}.}
    \label{fig:Raman_modes}
\end{figure}

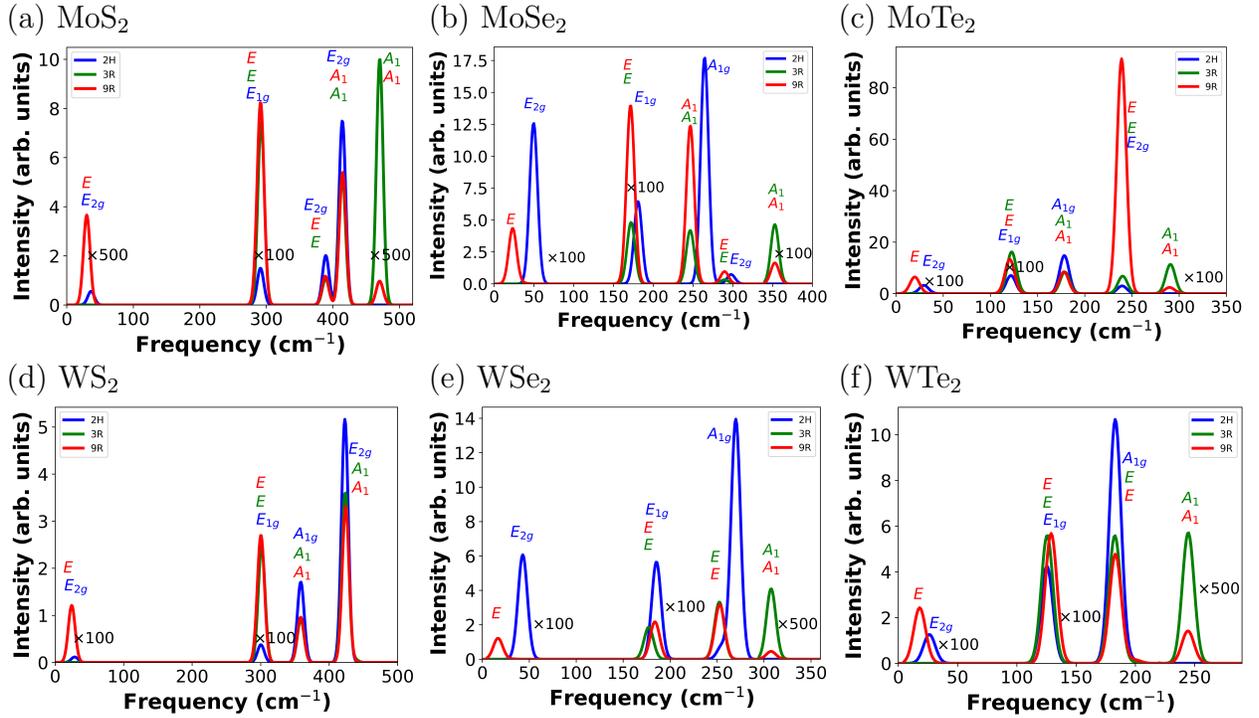
\begin{figure}[htbp]
			
			\centering{
			\begin{tikzpicture}
			\node [anchor=north west] (imgA) at (-0.15\linewidth,.58\linewidth){\includegraphics[width=0.335\linewidth]{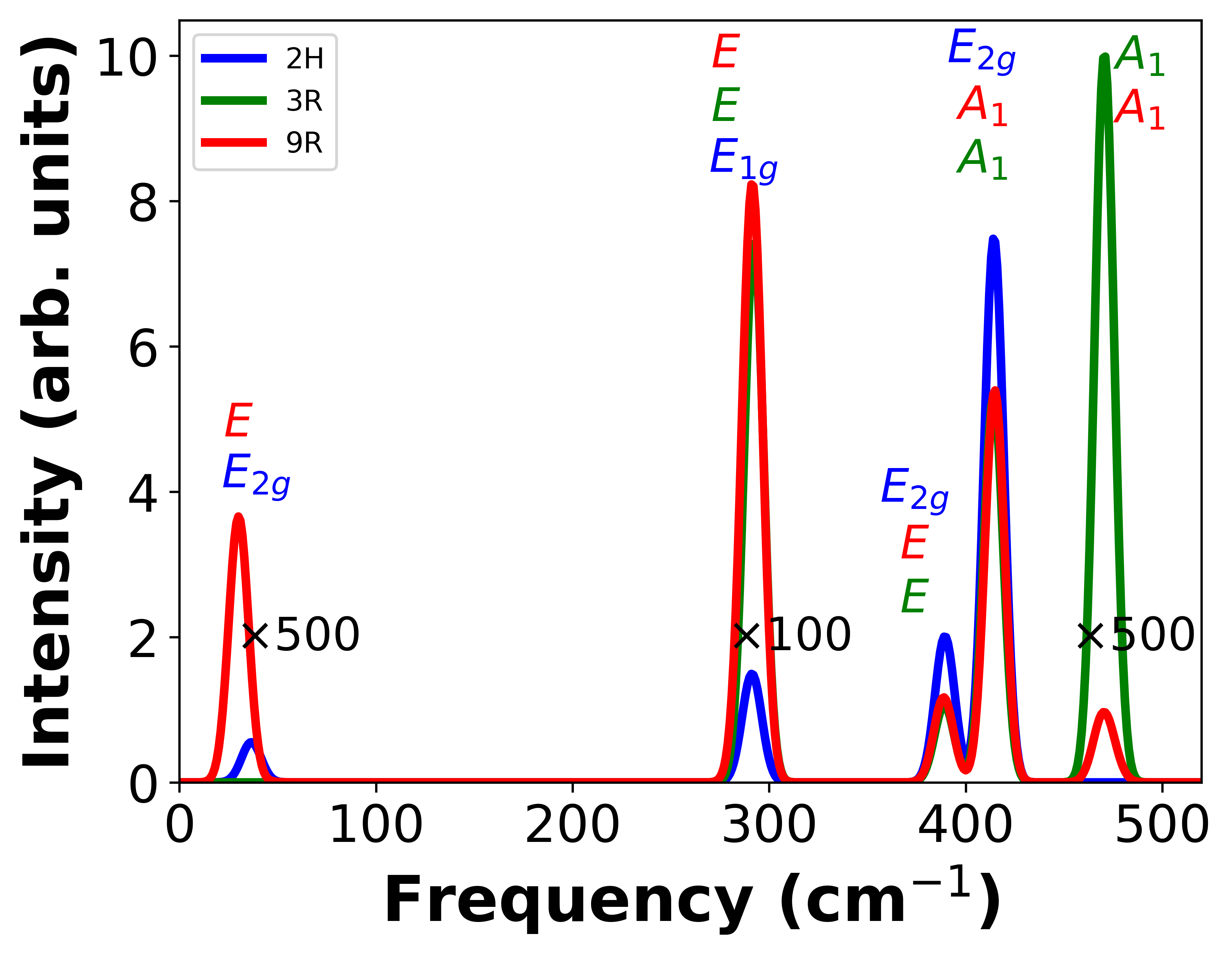}};
            \node [anchor=north west] (imgB) at (0.186\linewidth,.58\linewidth){\includegraphics[width=0.33\linewidth]{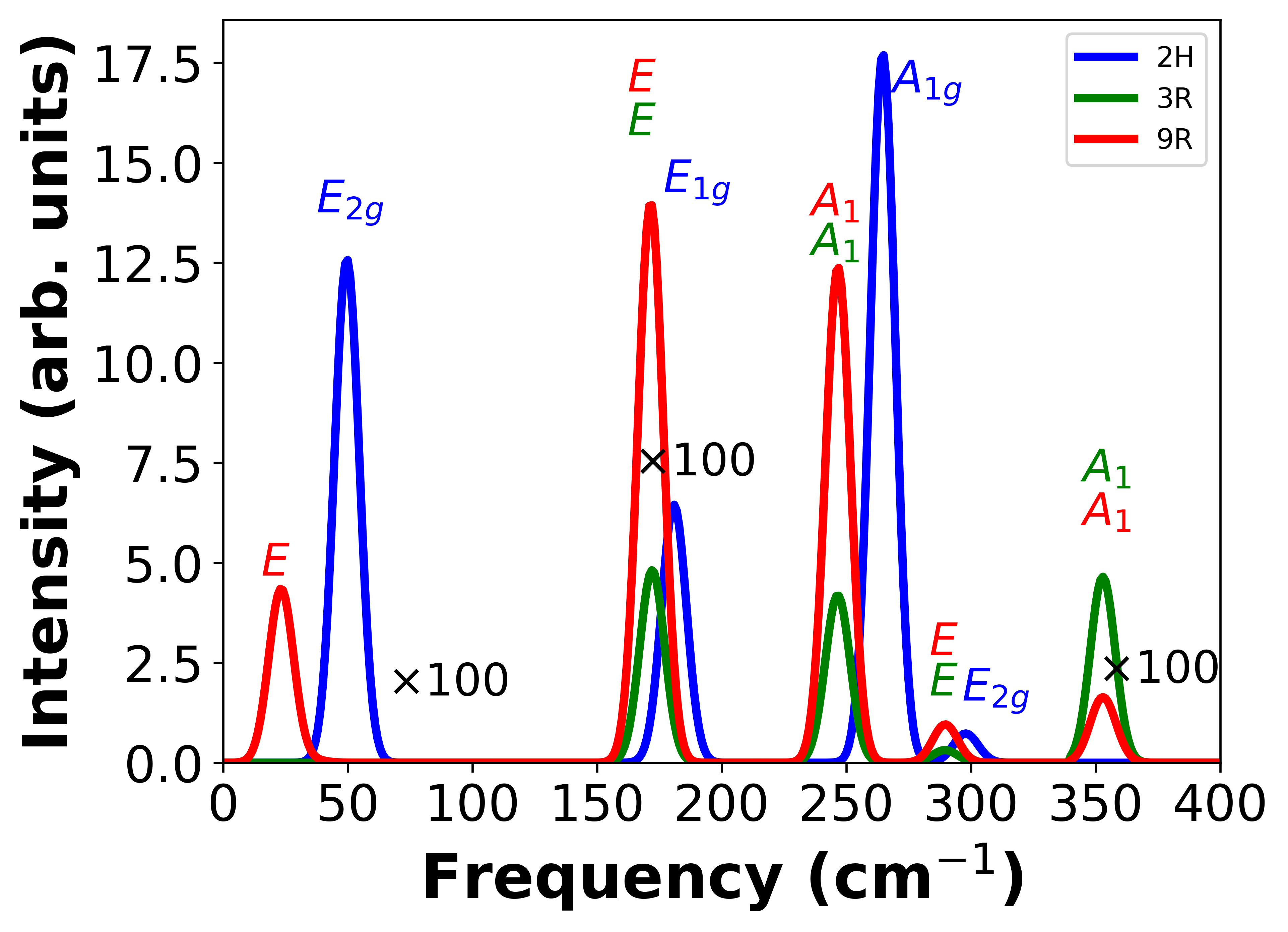}};
          \node [anchor=north west] (imgC) at (0.52\linewidth,.58\linewidth){\includegraphics[width=0.33\linewidth]{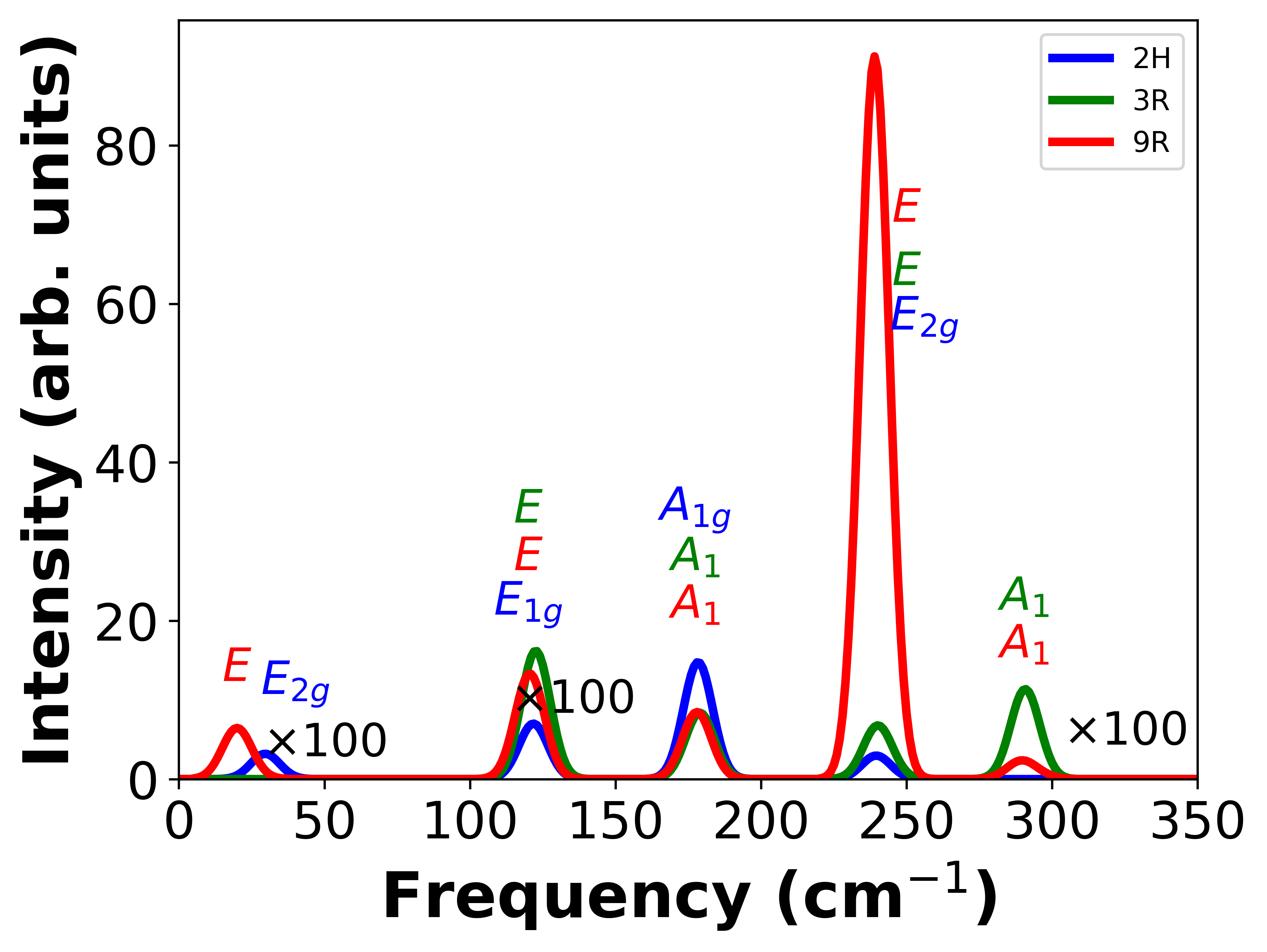}};
          
          	\node [anchor=north west] (imgD) at (-0.15\linewidth,.29\linewidth){\includegraphics[width=0.333\linewidth]{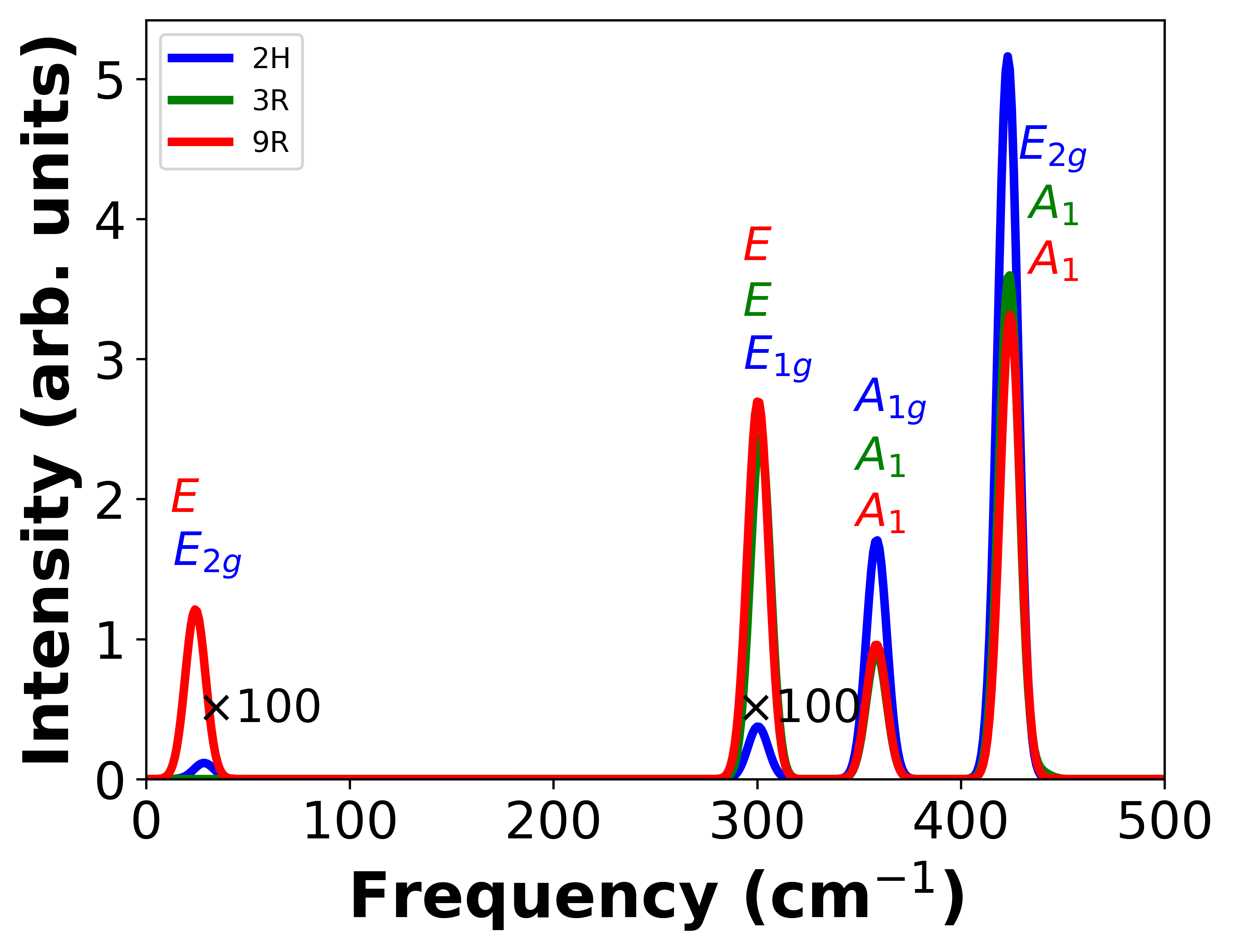}};
            \node [anchor=north west] (imgE) at (0.186\linewidth,.29\linewidth){\includegraphics[width=0.33\linewidth]{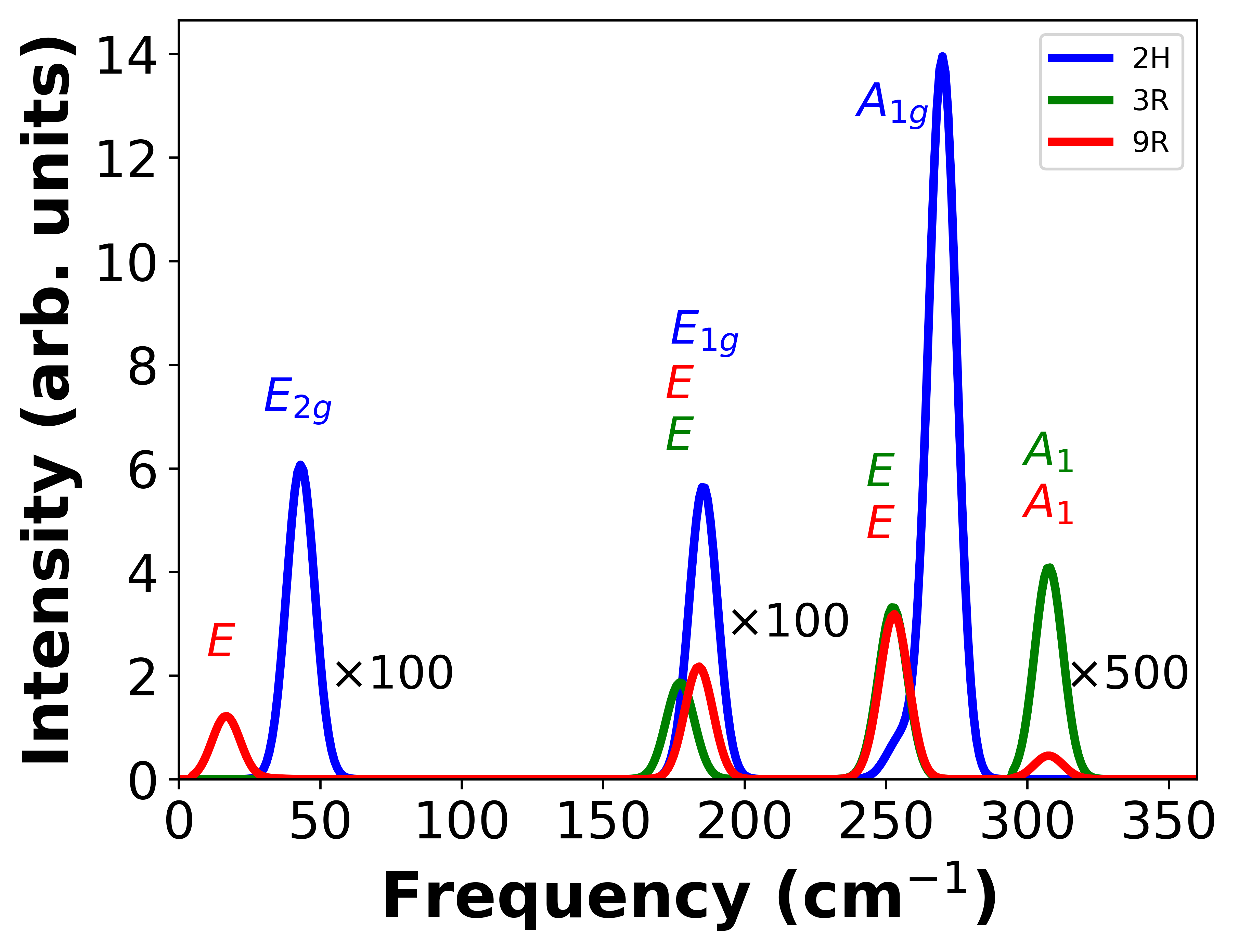}};
          \node [anchor=north west] (imgF) at (0.52\linewidth,.29\linewidth){\includegraphics[width=0.33\linewidth]{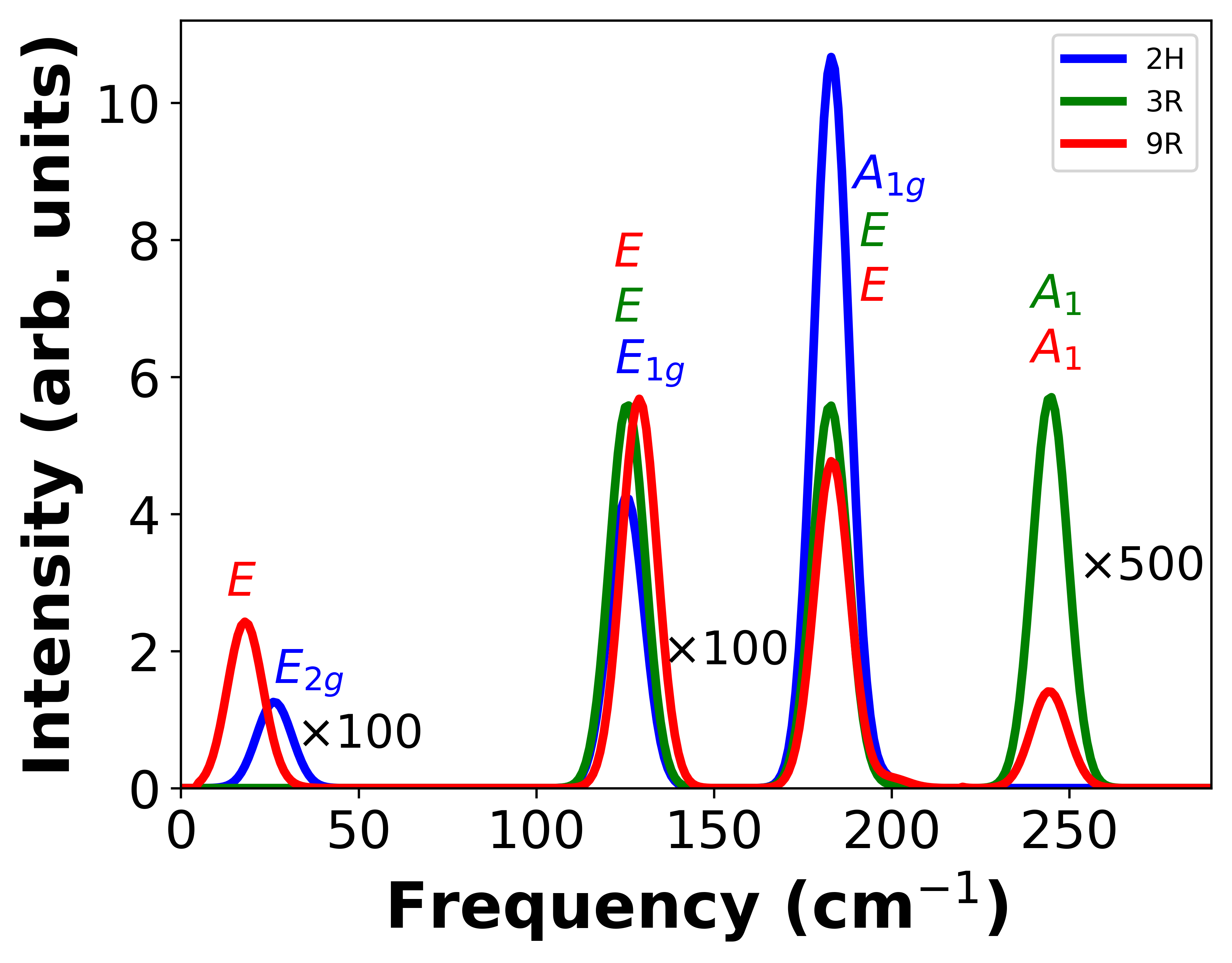}};
          
            \draw [anchor=north west] (-0.15\linewidth, .61\linewidth) node {(a) {\fontfamily{Arial}\selectfont \textbf{MoS$_2$}}};
            \draw [anchor=north west] (0.19\linewidth, .61\linewidth) node {(b) {\fontfamily{Arial}\selectfont \textbf{MoSe$_2$}}};
           \draw [anchor=north west] (0.52\linewidth, .61\linewidth) node {(c) {\fontfamily{Arial}\selectfont \textbf{MoTe$_2$}}};
            \draw [anchor=north west] (-0.15\linewidth, .32\linewidth) node {(d) {\fontfamily{Arial}\selectfont \textbf{WS$_2$}}};
            \draw [anchor=north west] (0.19\linewidth, .32\linewidth) node {(e) {\fontfamily{Arial}\selectfont \textbf{WSe$_2$}}};
           \draw [anchor=north west] (0.52\linewidth, .32\linewidth) node {(f) {\fontfamily{Arial}\selectfont \textbf{WTe$_2$}}};          
            \end{tikzpicture}
			}
			
			\caption{Calculated Raman spectra of different TMDs in 2H, 3R and 9R phases.  The highest frequency modes in WS$_2$  do exist (see Table \ref{tab:Raman_frequencies}) but are mixed to larger nearby peaks. In a few frequency ranges, we scaled the intensities (shown by a multiplicative factor for all phases in that range) for visualization purposes.   }
		\label{fig:Raman_9R}
			
		\end{figure}

We conducted a comprehensive characterization of the Raman spectra for all bulk phases across six different TMDs. Figure \ref{fig:Raman_9R} presents the calculated Raman spectra for these TMDs in all bulk phases, and detailed information regarding peak positions and the corresponding mode characteristics can be found in Table \ref{tab:Raman_frequencies}. Notably, both the 2H and 9R phases exhibit a distinct peak in the low-frequency regime (below 50 cm$^{-1}$), which is notably absent in the 3R phase. Conversely, both the 3R and 9R phases feature a high-frequency peak, with the specific frequency varying depending on the type of TMD. This high-frequency peak is not observed in the 2H phase. Based on these observations, we propose that it is possible to distinguish between the 9R, 3R, and 2H phases in Raman experiments by comparing the  Raman shifts at both the low- and high-frequency regimes. In the majority of cases, the Raman active modes in the 2H, 3R, and 9R phases exhibit similar vibration patterns. At the low-frequency regime, we observe layer vibrations in opposite directions in both the 2H and 9R phases as shown in \ref{fig:Raman_modes}. In the mid-frequency range, the Raman active modes are dominated by D (Dichalcogenides) vibrations, which occur either in-plane in parallel and anti-parallel directions or out-of-plane. The widest Raman active regime is characterized by TM (Transition Metal) and D atoms vibrating in opposite directions out-of-plane in 3R and 9R (see Fig. \ref{fig:Raman_modes}). These shared vibration patterns provide valuable insights into the structural and vibrational characteristics of these phases.

Additionally, the 9R phase, which shares the lack of inversion symmetry with the 3R phase, should, in principle, exhibit second-harmonic generation (SHG). On the other hand, the 2H phase lacks this property. In experiments, researchers have often relied on SHG to differentiate between the 2H and 3R phases, as shown in  \cite{piezo_experiement}. However, we argue that SHG alone may not be sufficient to distinguish between 2H and 3R, especially given the existence of the 9R phase. Instead, we suggest that SHG can help rule out the 2H phase, while Raman spectroscopy can provide confirmation between the 3R and 9R phases, particularly by examining the low-frequency Raman peaks. To facilitate experimental investigations with SHG, we have provided two significant ratios for distinguishing between the 3R and 9R phases, as outlined in Table \ref{tab:SHG}. Specifically, we have found that the $\chi_{zzz}^{(2)} $ ratio is approximately twice as large in the 3R phase compared to the 9R phase. However, the ratio of $\chi_{yyy}^{(2)} $ varies, with the 3R phase exhibiting greater SHG intensity for the Mo family and the 9R phase displaying higher SHG intensity for the W family. This insight can guide SHG experiments in determining the intensity ratio between the 3R and 9R phases. The tensor form of $\chi^{(2)} $ for $C_{3v}$ point group is given by \cite{SHG_tensor}:

\begin{equation}
  \overleftrightarrow \chi^{(2)} =  
   \begin{pmatrix}
   \begin{pmatrix}
       0 & -\chi_{222} & \chi_{131} \\
       -\chi_{222} & 0 & 0 \\
       \chi_{113} & 0  & 0 \\
   \end{pmatrix}
   \\
   \begin{pmatrix}
       -\chi_{222} & 0 & 0 \\
       0 & \chi_{222} & \chi_{131} \\
       0 & \chi_{113} & 0 \\       
   \end{pmatrix}
   \\
   \begin{pmatrix}
       \chi_{311} & 0 & 0 \\
       0 & \chi_{311} & 0 \\
       0 & 0 & \chi_{333}
   \end{pmatrix}

   \end{pmatrix}
   \label{SHG_matrix}
\end{equation}

\begin{table}[H]
    \centering
    \footnotesize
    \begin{tabular}{|c|c|c|c|}
    
    \hline
    TMDs & 2H & 3R & 9R \\
    \hline
    \hline
    \multirow{2}{*}{MoS$_2$} & 36.58 ($E_{2g}$), 291.3 ($E_{1g}$), & 292.2 ($E$), 389.1 ($E$), & 30.3 ($E$), 292.1 ($E$), \\
    & 389.4 ($E_{2g}$), 414.4 ($E_{2g}$) & 414.1 ($A_1$), 470.5 ($A_1$) & 389.8 ($E$), 414.7 ($A_1$), 469.4 ($A_1$) \\
    \hline
  \multirow{2}{*}{MoSe$_2$} & 49.6 ($E_{2g}$), 180.9 ($E_{1g}$), & 172.1 ($E$), 246.5 ($A_1$), & 23.3 ($E$), 169.8 ($E$),172 ($E$), \\
    & 264.6 ($A_{1g}$), 297.9 ($E_{2g}$) & 289.6 ($E$), 352.9 ($A_1$) &  246.7 ($A_1$), 289.7 ($E$), 352.3 ($A_1$) \\
    \hline
\multirow{2}{*}{MoTe$_2$} & 29.6 ($E_{2g}$), 121.8 ($E_{1g}$), & 122.5 ($E$), 178.9 ($A_1$), & 20.5 ($E$), 120.6 ($E$), \\
    & 178.3 ($A_{1g}$), 239.5 ($E_{2g}$) & 240.1 ($E$), 290.7 ($A_1$) & 177.9 ($A_1$), 239 ($E$), 289.2 ($A_1$) \\
    \hline
\multirow{2}{*}{WS$_2$} & 28.4 ($E_{2g}$), 300.4 ($E_{1g}$), & 301.5 ($E$), 358.5 ($E$), & 24.1 ($E$), 298.7 ($E$), 301.4 ($E$), \\
    & 358.6 ($E_{2g}$), 423 ($A_{1g}$) & 423.6 ($A_1$), 438 ($A_1$) & 358.5 ($E$), 424.2 ($A_1$), 436.8 ($A_1$) \\
    \hline
\multirow{2}{*}{WSe$_2$} & 43 ($E_{2g}$), 185.4 ($E_{1g}$), & 177.2 ($E$), 248.8 ($E$), & 33.7 ($A_1$), 184.3 ($E$), \\
    & 255.1 ($E_{2g}$), 269.9 ($A_{1g}$) & 252.7 ($A_1$), 307.5 ($A_1$) & 248.9 ($E$), 253.3 ($A_1$), 306.8 ($A_1$) \\
    \hline
\multirow{2}{*}{WTe$_2$} & 26.3 ($E_{2g}$), 125.4 ($E_{1g}$), & 125.6 ($E$), 182.7 ($A_1$), & 18 ($A_1$), 127.3 ($E$),129.8 ($E$), \\
    & 183 ($A_{1g}$), 198.9 ($E_{2g}$) & 199.2 ($E$), 244.6 ($A_1$) & 183.8 ($A_1$), 199.6 ($E$), 243.8 ($A_1$) \\

    \hline
    \end{tabular}
    \caption{Raman active modes (in cm$^{-1}$) and their corresponding character are given in parenthesis.}
    \label{tab:Raman_frequencies}
\end{table}

\begin{table}[htbp]
    \centering
    \begin{tabular}{|c|r|r|}
    \hline
  TMDs & $\frac{^{9R}\chi_{yyy}^{(2)}} {^{3R}\chi_{yyy}^{(2)}}$ & $\frac{^{9R}\chi_{zzz}^{(2)}} {^{3R}\chi_{zzz}^{(2)}}$ \\
\hline
\hline
     MoS$_2$& 0.75 & 0.37  \\
     \hline
          MoSe$_2$& 0.06 & 0.40  \\
     \hline
          MoTe$_2$& 0.46 & 0.37 \\
     \hline
          WS$_2$& 3.70 & 0.37  \\
     \hline
          WSe$_2$& 7.14 & 0.43  \\
     \hline
          WTe$_2$& 8.33 & 0.45\\
     \hline
     
    \end{tabular}
    \caption{SHG ratios of 3R and 9R.}
    \label{tab:SHG}
\end{table}

\begin{figure}[htbp]
			\centering{
			\begin{tikzpicture}
			\node [anchor=north west] (imgA) at (-0.15\linewidth,.58\linewidth){\includegraphics[width=0.335\linewidth]{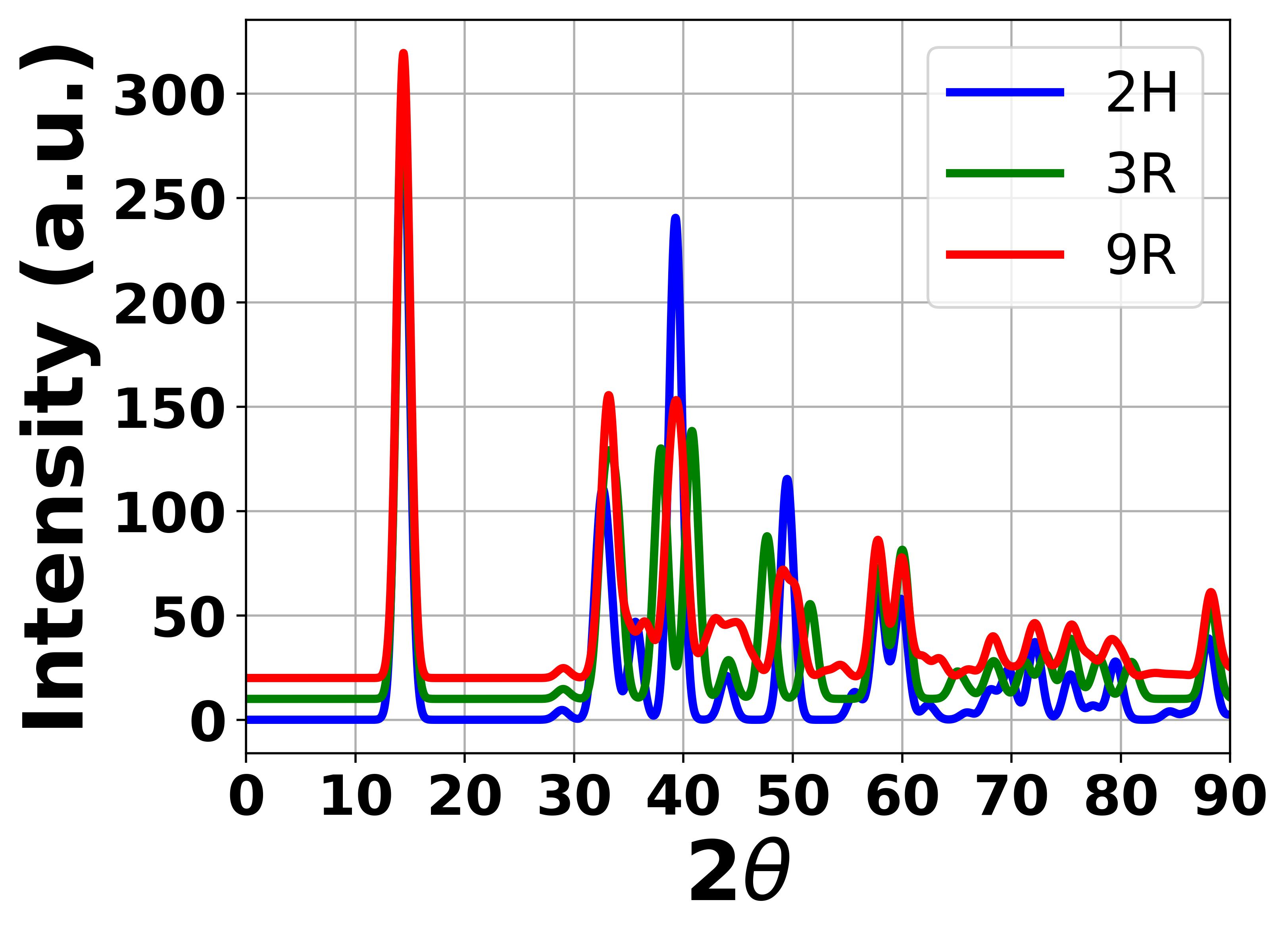}};
            \node [anchor=north west] (imgB) at (0.186\linewidth,.58\linewidth){\includegraphics[width=0.33\linewidth]{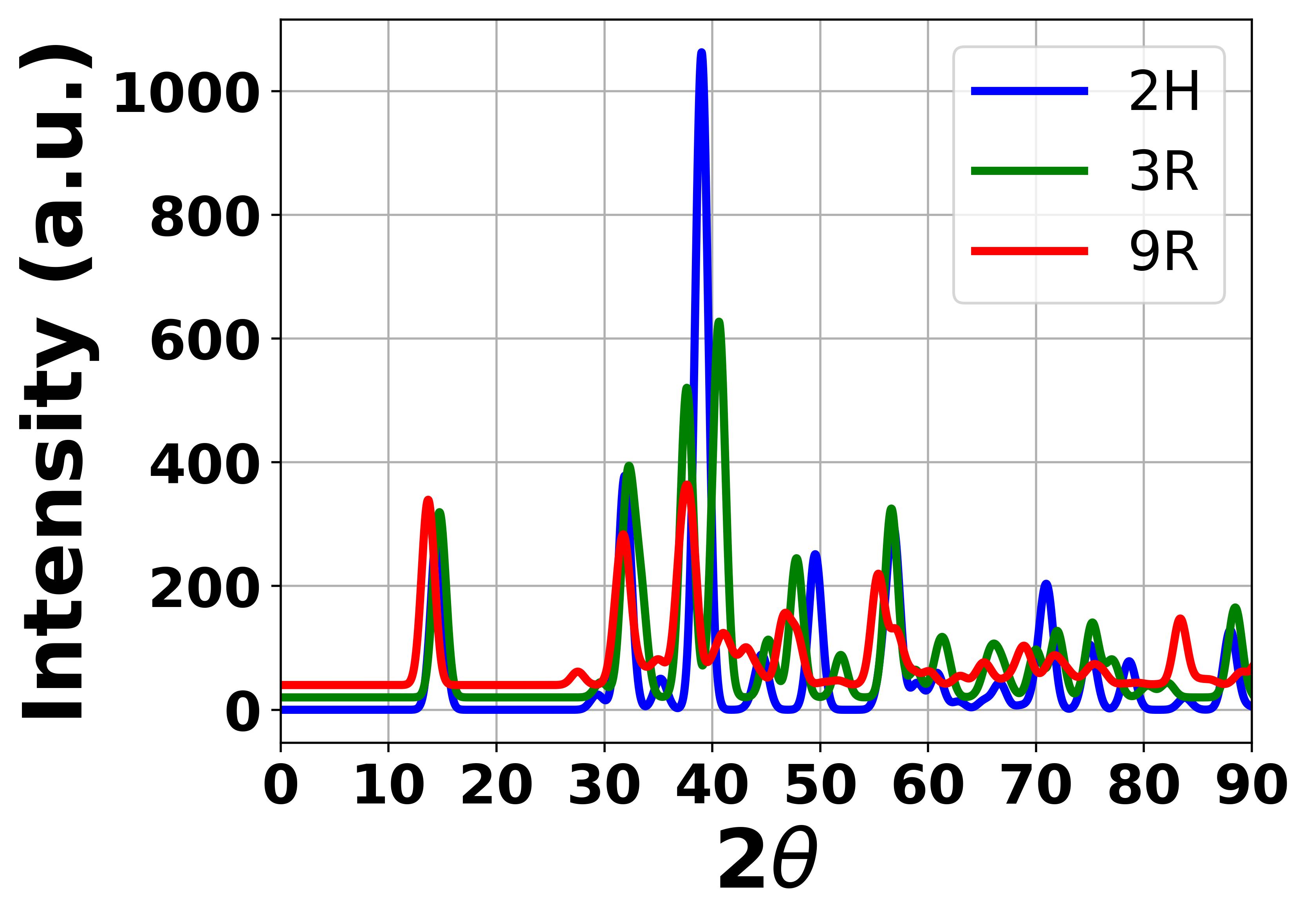}};
          \node [anchor=north west] (imgC) at (0.52\linewidth,.58\linewidth){\includegraphics[width=0.33\linewidth]{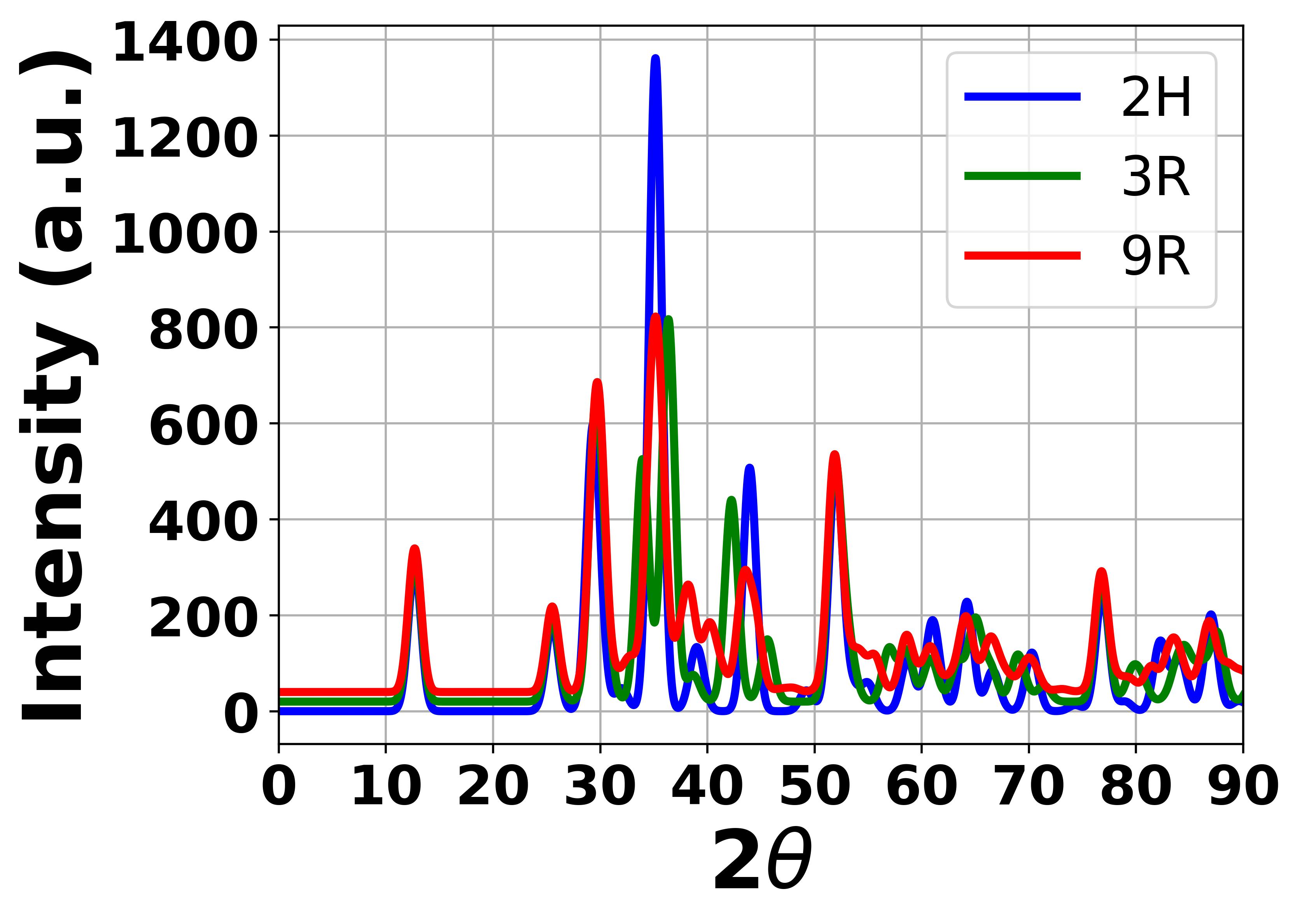}};
          
          	\node [anchor=north west] (imgD) at (-0.15\linewidth,.30\linewidth){\includegraphics[width=0.333\linewidth]{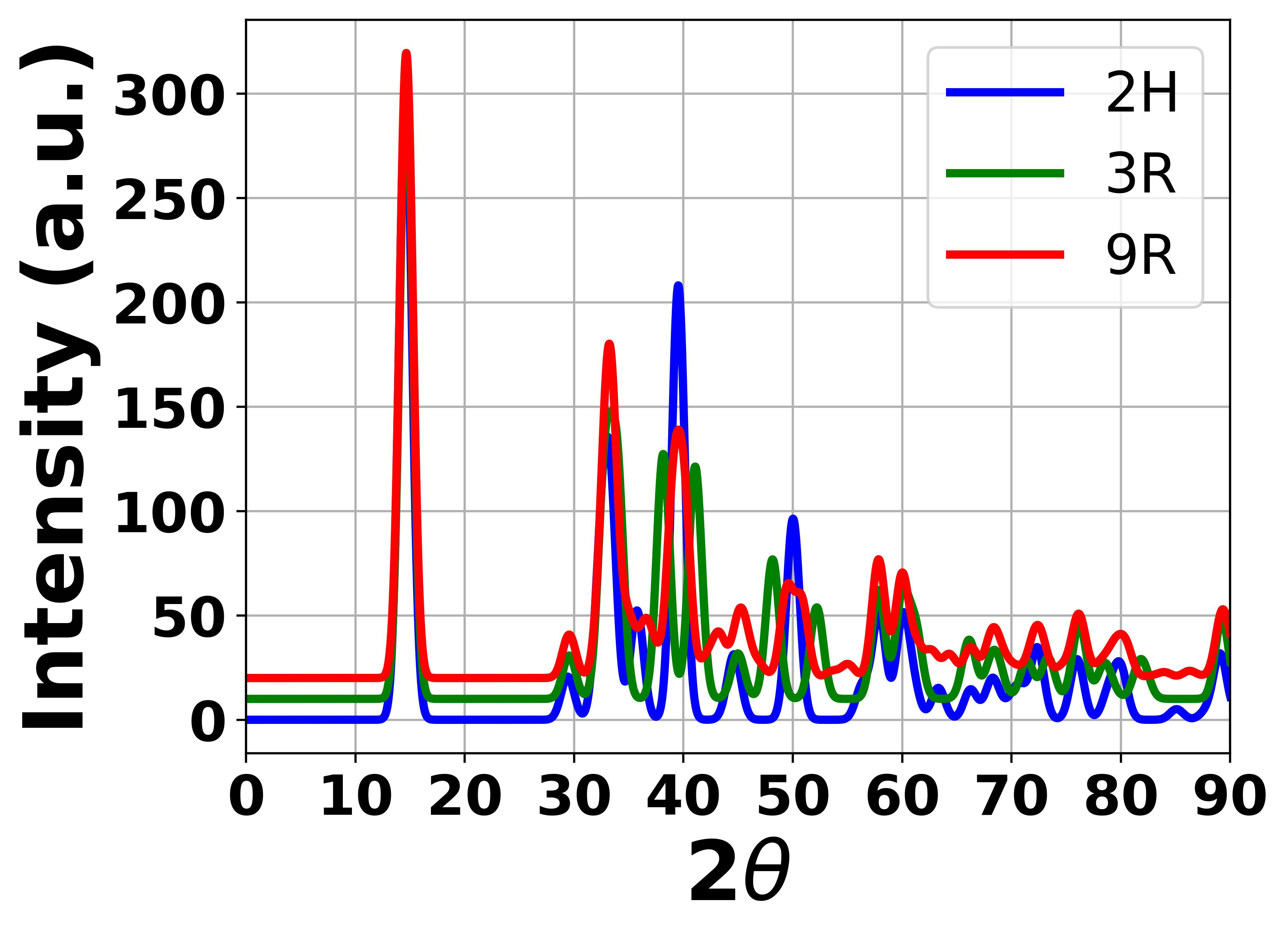}};
            \node [anchor=north west] (imgE) at (0.186\linewidth,.30\linewidth){\includegraphics[width=0.33\linewidth]{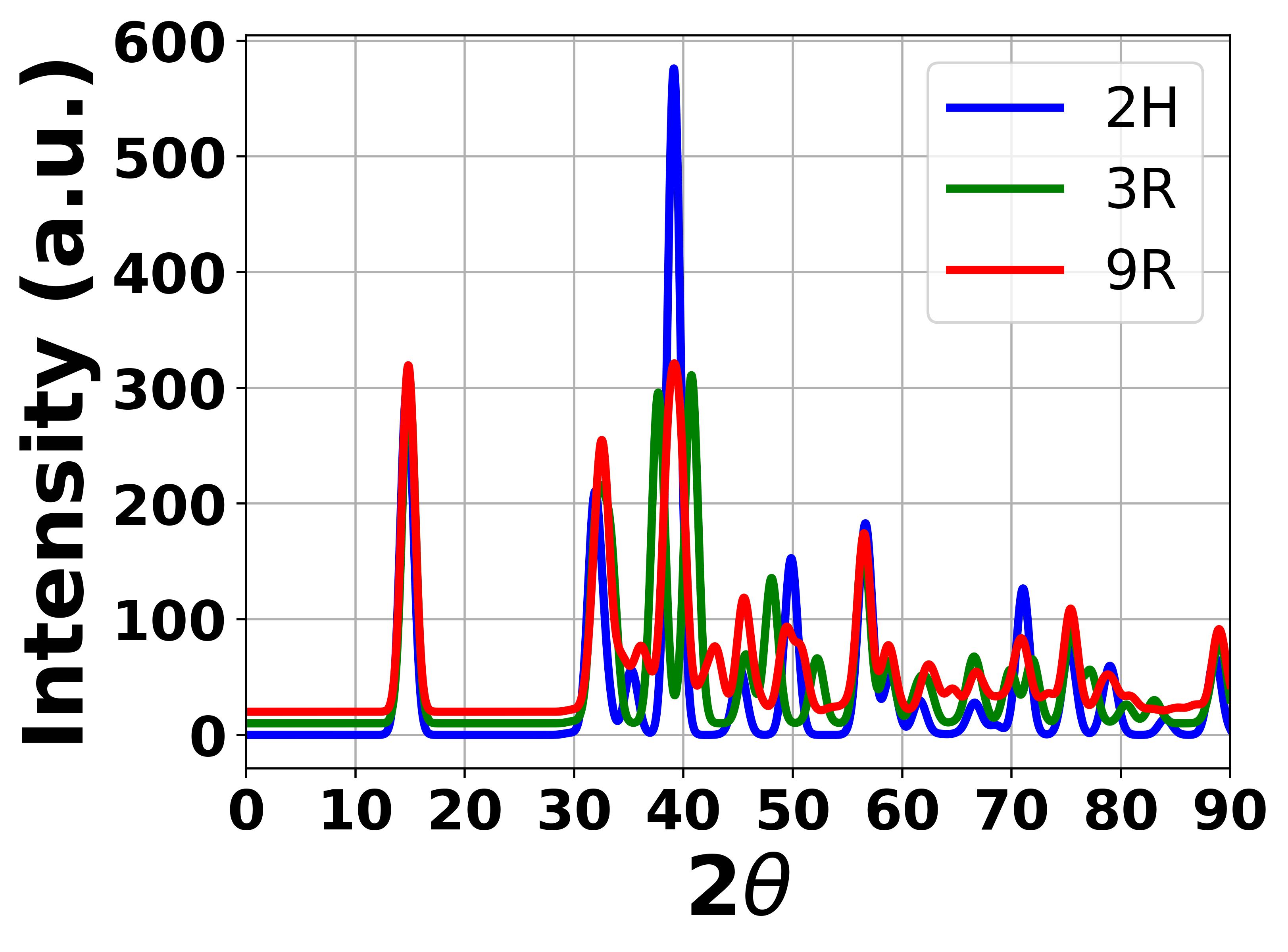}};
          \node [anchor=north west] (imgF) at (0.52\linewidth,.30\linewidth){\includegraphics[width=0.33\linewidth]{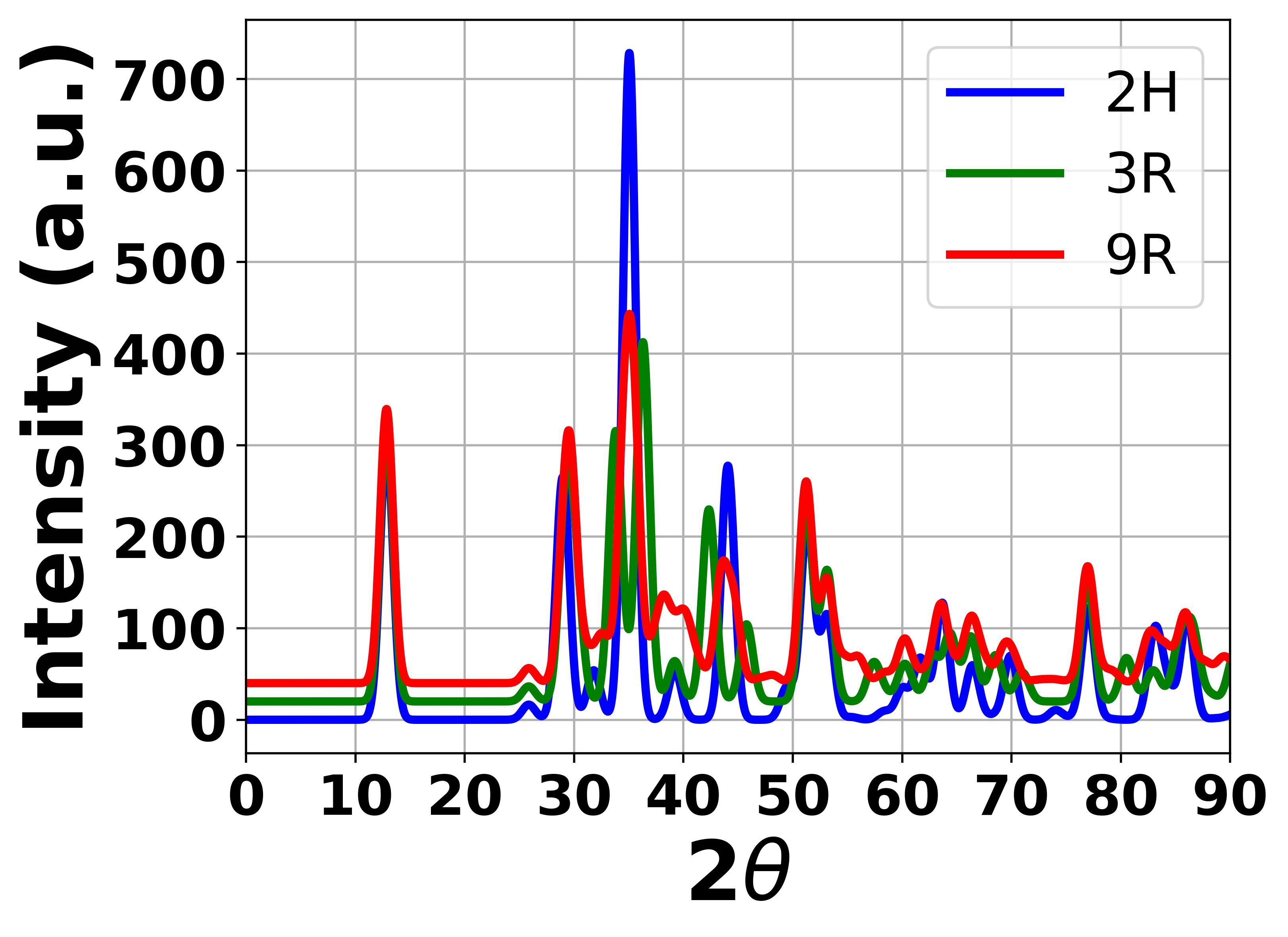}};
          
            \draw [anchor=north west] (-0.15\linewidth, .61\linewidth) node {(a) {\fontfamily{Arial}\selectfont \textbf{MoS$_2$}}};
            \draw [anchor=north west] (0.19\linewidth, .61\linewidth) node {(b) {\fontfamily{Arial}\selectfont \textbf{MoSe$_2$}}};
           \draw [anchor=north west] (0.52\linewidth, .61\linewidth) node {(c) {\fontfamily{Arial}\selectfont \textbf{MoTe$_2$}}};
            \draw [anchor=north west] (-0.15\linewidth, .33\linewidth) node {(d) {\fontfamily{Arial}\selectfont \textbf{WS$_2$}}};
            \draw [anchor=north west] (0.19\linewidth, .33\linewidth) node {(e) {\fontfamily{Arial}\selectfont \textbf{WSe$_2$}}};
           \draw [anchor=north west] (0.52\linewidth, .33\linewidth) node {(f) {\fontfamily{Arial}\selectfont \textbf{WTe$_2$}}};          
            \end{tikzpicture}
			}
			
			\caption{Calculated diffraction patterns of different TMDs in 2H, 3R and 9R phases. The peaks of 3R and 9R are shifted vertically for clarity.}
		\label{fig:diffraction_patterns}
			
		\end{figure}
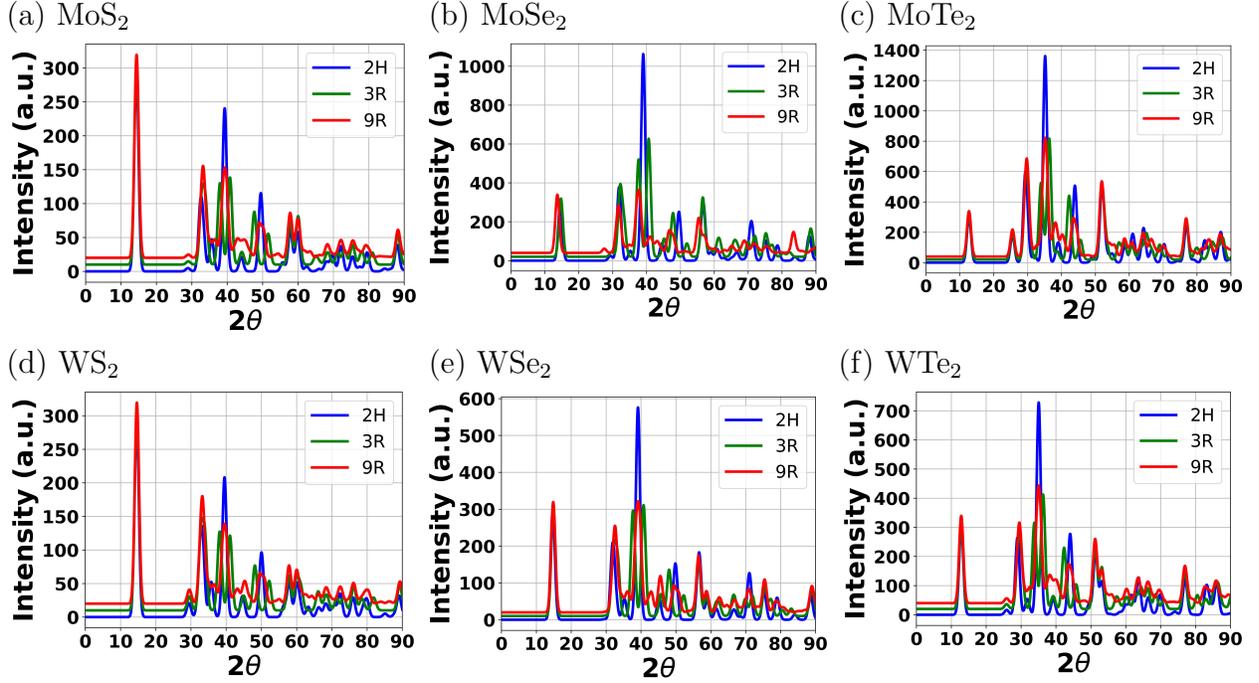

We also performed calculations of the powder diffraction patterns on the TMDs based on VESTA using methodology described by  \cite{izumi2007}, with the aim of providing guidance to experimentalists for detecting the presence of the 9R phase. A crucial distinction becomes evident in the $2\theta$ at  $39^{\circ}$ for 2H and 9R, and $38^{\circ}$ and $40^{\circ}$ for 3R, for all except WTe$_2$ which has similar feature at $35^{\circ}$ for 2H and 9R, and $34^{\circ}$ and $36^{\circ}$ for 3R, as illustrated in Figure \ref{fig:diffraction_patterns}. The crystallographic planes $hkl$ corresponding to these angles are $(013)$ for 2H, $(554)$ for 9R, and $(01\overline{4})$ and $(\overline{1}1\overline{5})$ for two adjacent peaks in 3R. These characteristic diffraction patterns could also help in the identification or differentiation of structural phases in TMDs.

\subsection{Important features}
We explored some properties, such as electronic band structure, the optoelectric tensor for SHG but found it similar to the 3R phase. We found two properties that are better than the 3R phase and is discussed below.

\subsubsection{Piezoelectricity}
Piezoelectric materials find application in various significant devices, including microphones, medical imaging tools, and sensors. \cite{piezo1,piezo2} Notably, recent advancements have showcased the utility of piezopotentials generated by piezoelectricity as a gate voltage for manipulating the electronic band gap in piezoelectric semiconductors. This development has brought a novel research domain known as ``piezotronics". \cite{piezo3} In this context, 2D semiconductors are particularly promising materials, given their ability to withstand the substantial deformations encountered in piezoelectric applications.
The piezoelectric properties can be calculated by evaluating the change in the electrical polarization in response to the applied strain and can be expressed as:
\begin{equation}
e_{ijk}=\frac{dP_{i}}{d\epsilon_{jk}}
\end{equation}
Here, $P_i$ is the electrical polarization along $i^{th}$ direction due to strain along $j,k$ direction where $i,j,k$ $\in$ {1,2,3}, with 1,2,3 corresponding to $x$, $y$, and $z$ directions, respectively. Likewise, piezoelectric strain coefficients $d_{ij}$ can be obtained from the stress coefficients and elastic constants $C_{ij}$ by:
\begin{equation}
\label{}
e_{ik}= \sum_{j=1}^{6} d_{ij} C_{jk}
\end{equation}

The $d_{ij}$ matrix tensor for 3R or 9R belonging to $3m$ space group and $C_{3v}$ point group has only 4 non-zero elements \cite{Konabe} ($d_{15}, d_{22}, d_{31},$ and $d_{33}$) resulting in:
\begin{equation}
\begin{pmatrix}
0 & 0 & 0 & 0 & d_{15} & -d_{22} \\
-d_{22} & d_{22} &  0 & d_{15} & 0 & 0 \\
d_{31} & d_{31} & d_{33} & 0 & 0& 0 \\

\end{pmatrix}
\end{equation}

\begin{figure}[H]
    \centering
    \includegraphics[scale=0.45]{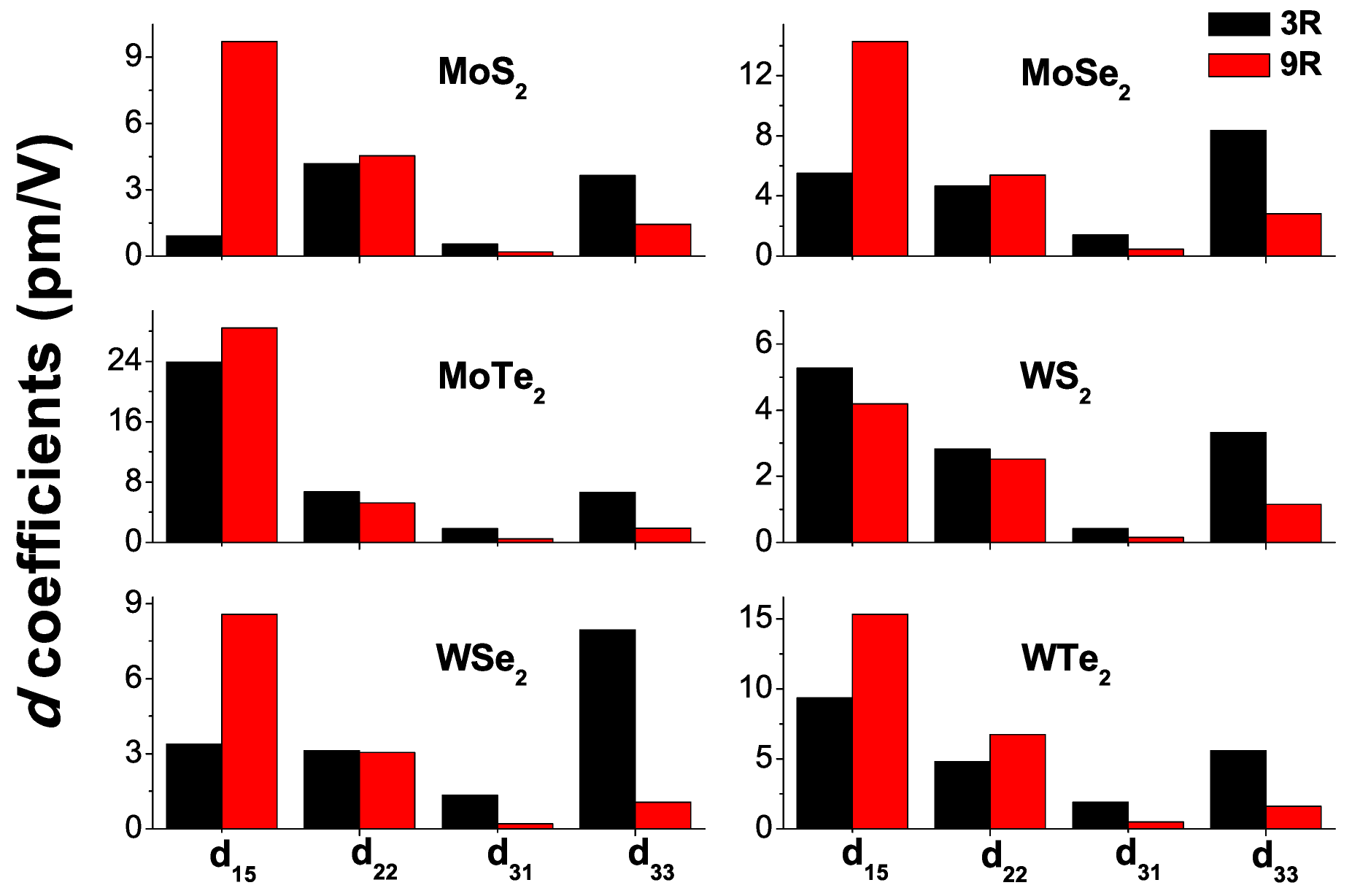}
    \caption{Piezoelectric coefficients on various TMDs. }
    \label{fig:piezo_comparisonl}
\end{figure}

Considering the data presented in Table \ref{Tab:piezo tensor} and Fig. \ref{fig:piezo_comparisonl}, it's evident that $d_{15}$ and $d_{22}$ tend to exhibit greater values in 9R compared to 3R structures. There are not many theoretical or experimental works on 3R phase of TMDs. One theoretical work   \cite{Konabe} reports first principle calculation on $e_{ij}$ and $d_{ij}$ coefficients for  MoS$_2$, MoSe$_2$, WS$_2$ and  WSe$_2$. In comparison with  \cite{Konabe}, $e_{ij}$ are similar in magnitudes but some of the  $d_{ij}$ coefficients are significantly smaller or larger than what we calculated. For example, we calculated $d_{33}$ for 3R MoS$_2$ as 3.65 pm/V and they reported 0.27 pm/V. Also, a recent experiment on 3R MoS$_2$ flakes with the thickness from 4 to 90 nm, or $\approx$ 6 to $\approx$ 128 layers showed $d_{33}$ in a range 0.7 
$\pm$ 0.2 to 1.5 $\pm$ 0.2 pm/V. \cite{piezo_experiement} In  \cite{Konabe}, some of the elastic coefficients $C_{14}$, $C_{33}$ and $C_{44}$ are significantly higher than this work or other theoretical calculations. \cite{3R_review} This might have led to discrepency in some of the $d_{ij}$ coefficients. The $d_{15}$ and $d_{22}$ are also significantly larger in magnitude than their monolayer counterparts computed from clamped ion method ($d_{11}$ ranges from 1.88- 4.33 pm/V) and relaxed-ion method ($d_{11}$ ranges from 2.19- 9.13 pm/V) among the discussed TMDs. \cite{monolayer_piezo} Also, $d_{15}$ and $d_{22}$ in 3R and 9R are greater than other known bulk piezoelectric materials such as $\alpha$-quartz ($d_{11}=2.3$   pm/V), \cite{PhysRev.110.1060} wurtzite GaN ($d_{33} = 3.1$  pm/V), and wurtzite AlN ($d_{33} = 5.1$ pm/V). \cite{bulk_piezo_GaN10.1063/1.1317244}
Depending on the specific application requirements, we can select the most suitable material by prioritizing the one with the highest piezoelectric coefficients. 

\subsubsection{Band-splitting}
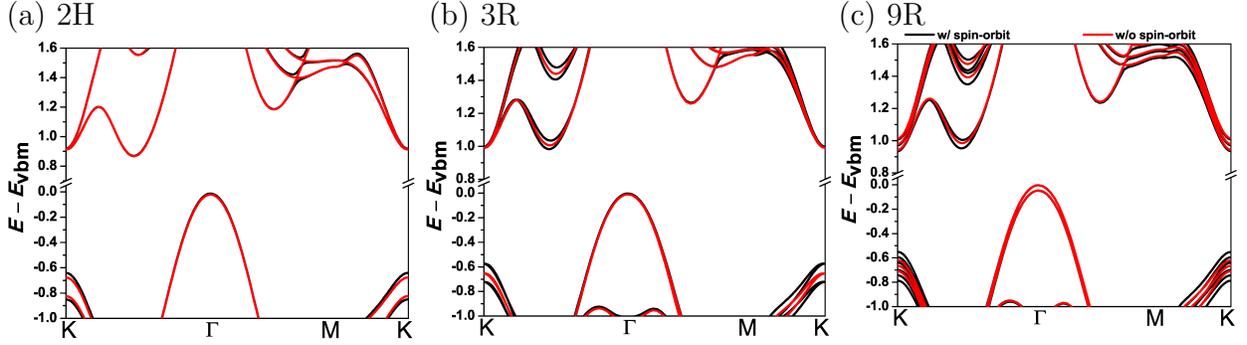
\begin{figure}[H]
			\centering{
			\begin{tikzpicture}
			\node [anchor=north west] (imgA) at (-0.15\linewidth,.58\linewidth){\includegraphics[width=0.33\linewidth]{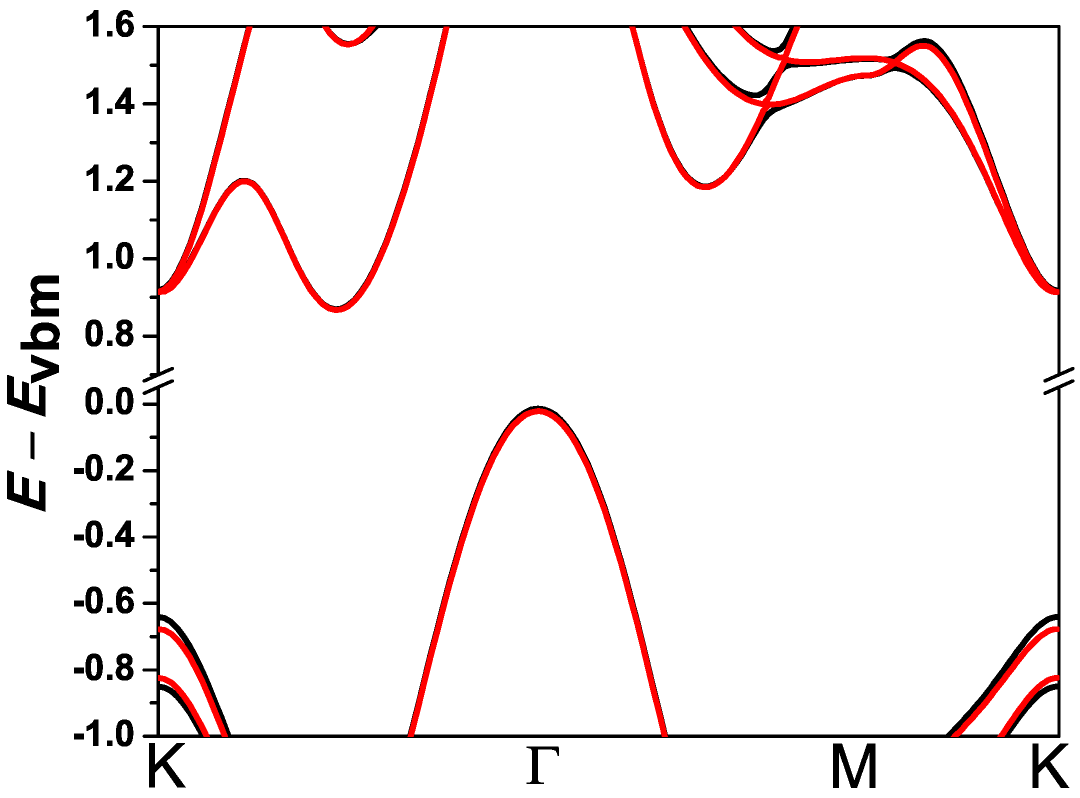}};
            \node [anchor=north west] (imgB) at (0.186\linewidth,.58\linewidth){\includegraphics[width=0.33\linewidth]{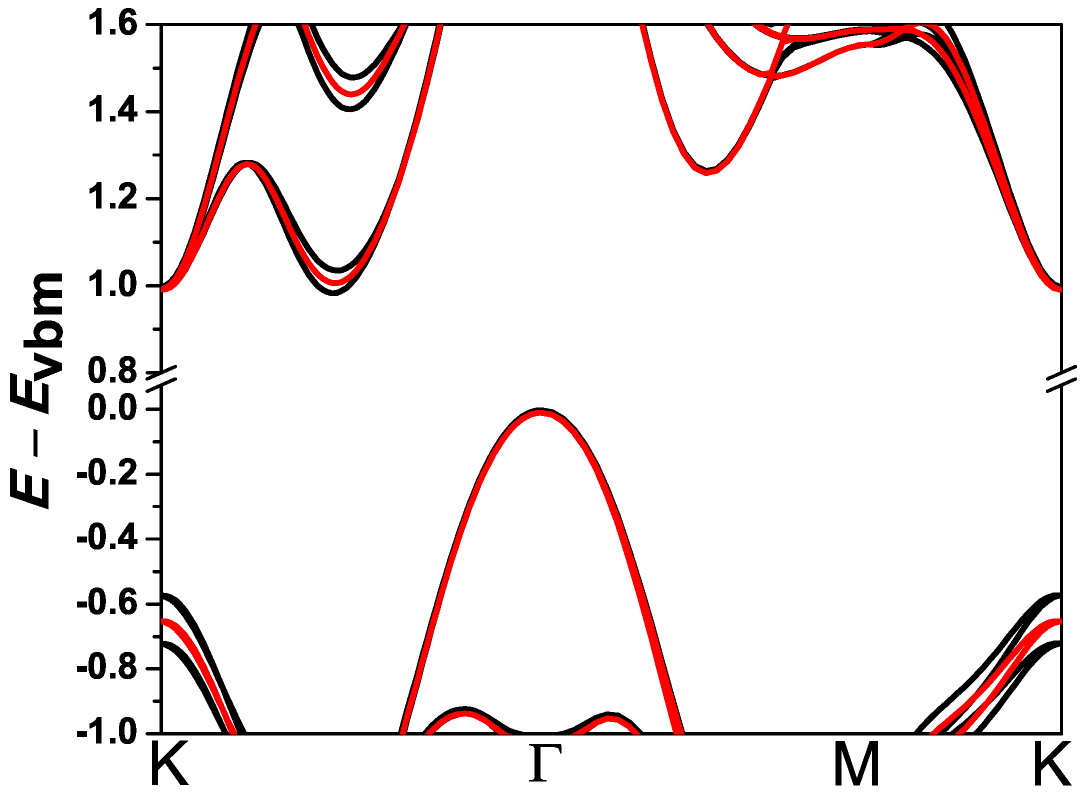}};
          \node [anchor=north west] (imgC) at (0.52\linewidth,.5895\linewidth){\includegraphics[width=0.33\linewidth]{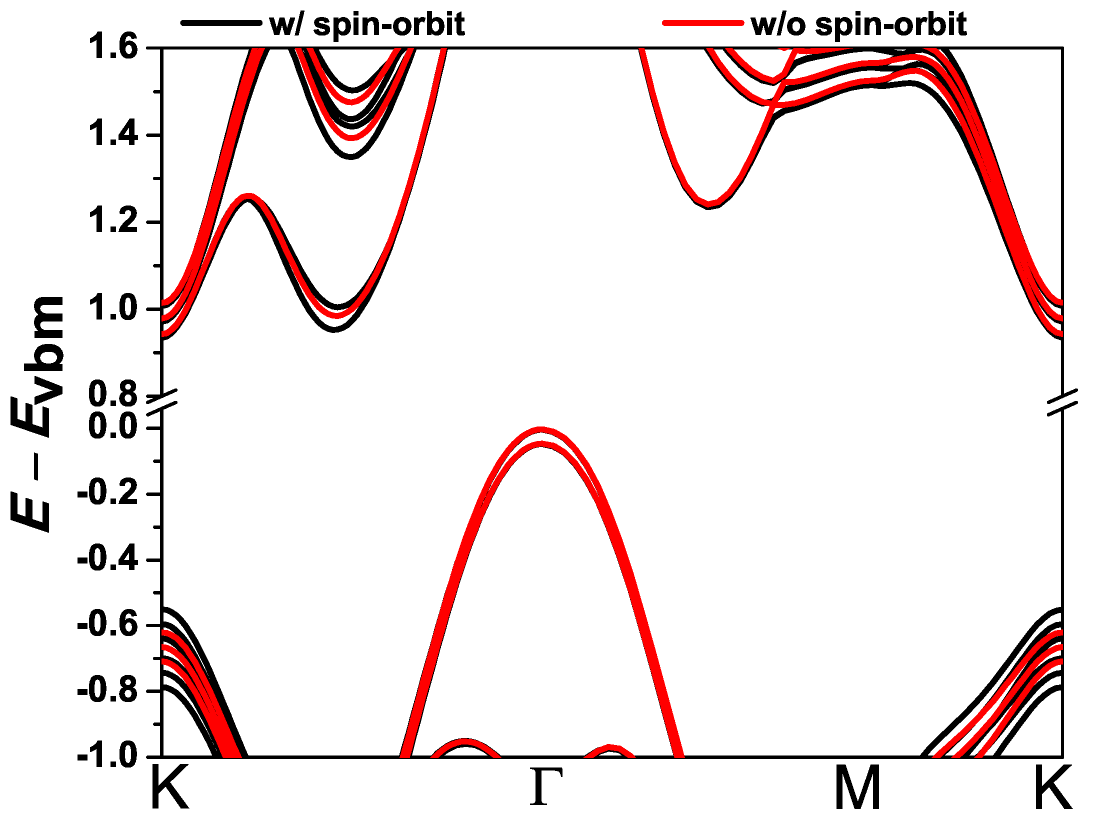}};

            \draw [anchor=north west] (-0.15\linewidth, .61\linewidth) node {(a) {\fontfamily{Arial}\selectfont \textbf{2H}}};
            \draw [anchor=north west] (0.19\linewidth, .61\linewidth) node {(b) {\fontfamily{Arial}\selectfont \textbf{3R}}};
           \draw [anchor=north west] (0.52\linewidth, .61\linewidth) node {(c) {\fontfamily{Arial}\selectfont \textbf{9R}}};
      
            \end{tikzpicture}
			}
			
			\caption{Electronic bandstructure with and without spin-orbit coupling in MoS$_2$.}
		\label{fig:9r_bandstructure_spin_orbit}
			
		\end{figure}

\begin{figure}[H]
			
			\centering{
			\begin{tikzpicture}
			\node [anchor=north west] (imgA) at (-0.15\linewidth,.58\linewidth){\includegraphics[width=0.33\linewidth]{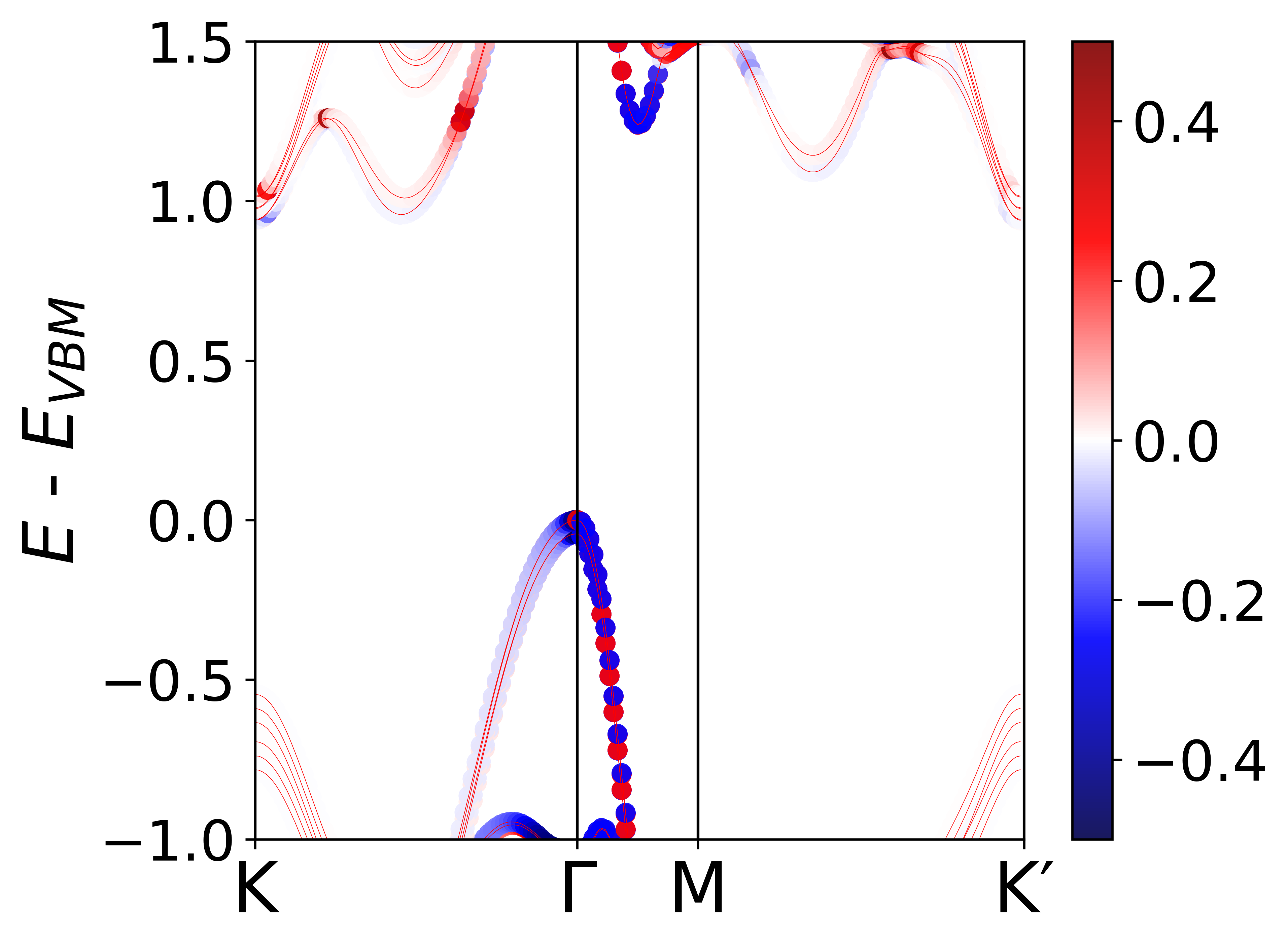}};
            \node [anchor=north west] (imgB) at (0.186\linewidth,.58\linewidth){\includegraphics[width=0.33\linewidth]{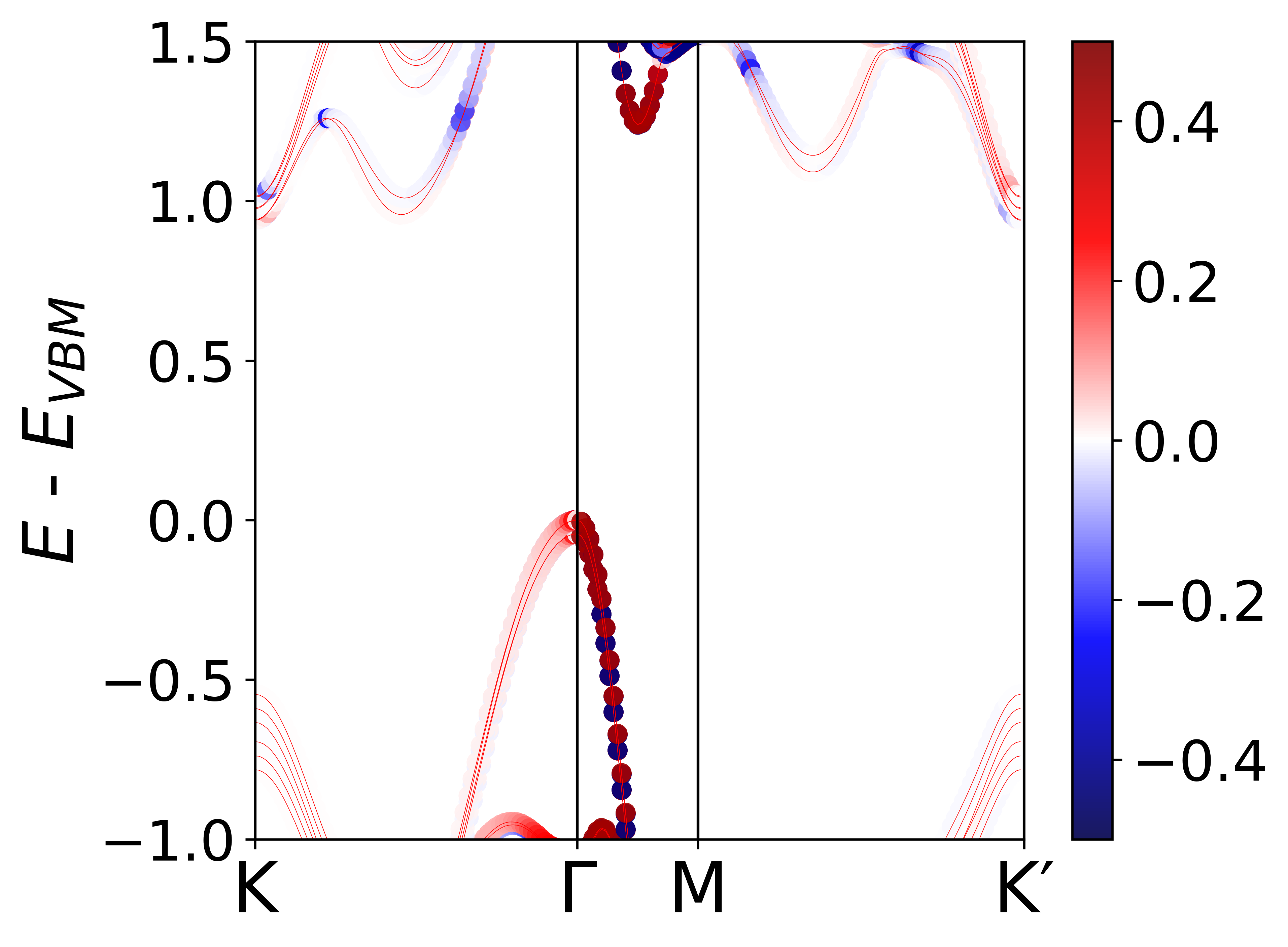}};
          \node [anchor=north west] (imgC) at (0.52\linewidth,.5895\linewidth){\includegraphics[width=0.33\linewidth]{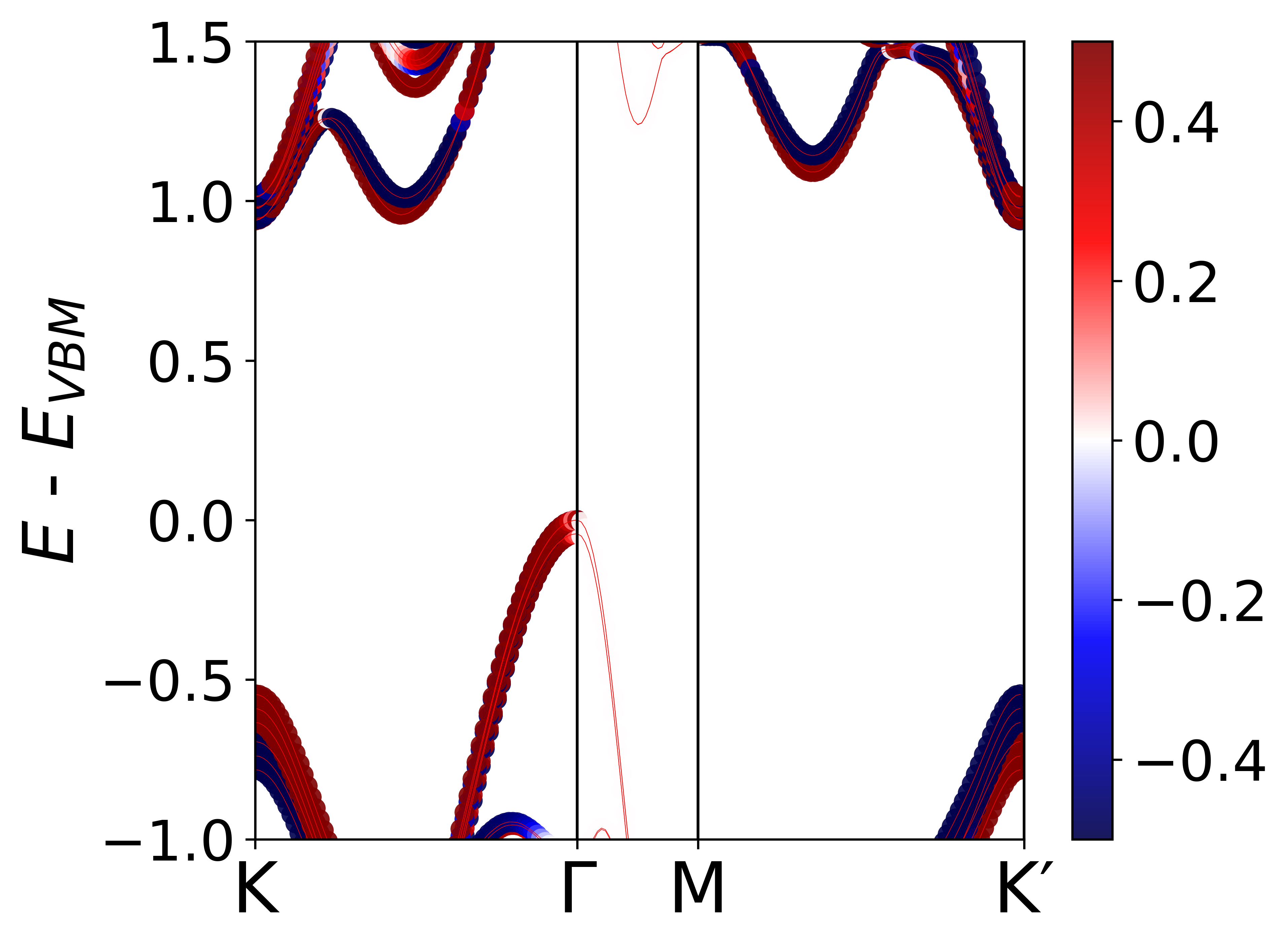}};

            \draw [anchor=north west] (-0.15\linewidth, .61\linewidth) node {(a) {\fontfamily{Arial}\selectfont \textbf{$<S_x>$}}};
            \draw [anchor=north west] (0.19\linewidth, .61\linewidth) node {(b) {\fontfamily{Arial}\selectfont \textbf{$<S_y>$}}};
           \draw [anchor=north west] (0.52\linewidth, .61\linewidth) node {(c) {\fontfamily{Arial}\selectfont \textbf{$<S_z>$}}};
      
            \end{tikzpicture}
			}
			
			\caption{Electronic bandstructure of 9R MoS$_2$ showing spin projections.}
		\label{fig:9r_bandstructure_spin_channels}
			
		\end{figure}

The monolayers TMDs lacks inversion symmetry and because of this symmetry breaking and stronger spin-orbit splittings, the two degenerate $K$ and $K'$ follows different optical selection rule due to opposite spin degenerate states. \cite{spintronics_appl3} Such spin-valley locking phenomenon has been studied for valleytronics devices such as qubits in quantum computing, low-power transisters, circularly polarized light emitters, polarization detectors, etc. \cite{valleytronics_review, spintronics_appl1, spintronics_appl2} In bulk crystals, this band separation primarily arises from a combination of spin-orbit coupling and interlayer coupling effects. \cite{spin_orb_tmd} Despite the relatively weaker interlayer interactions in TMDs, they can influence the states near the band edges, particularly affecting the valence band maximum and conduction band minimum, which are sensitive to interlayer coupling. \cite{interlayer_couplings} Previous literature has studied  different multilayer MoS$_2$ with different stacking sequences \cite{3R_MoS2_spin_modulation}. In our calculations, we observed band splitting in 2H, 3R and 9R bulk structures within the conduction and valence bands. Prior research on few-layer 3R-MoS$_2$ has demonstrated that the band splitting at the conduction band minimum is primarily dominated by layer coupling, with spin-orbit coupling perturbing this behavior. \cite{3R_MoS2_spin_modulation} In our calculation in bulk structures, we find no splitting in the  CBM of 2H, and slightly equal splitting due to spin orbit coupling in 3R and 9R.  In the same paper   \cite{3R_MoS2_spin_modulation}, the valence band splitting at $K$ is mainly contributed by spin-orbit  but we find mix contribution in band splitting at valence band at $K$ in bulk structures. The band splitting at $K$ of 2H is mainly due to layer coupling  and inclusion of spin-orbit has little contribution. In 3R at VB $K$ point, the splitting is only due to spin-orbit coupling whereas at same point in 9R, we find the combined contribution of layer coupling and spin-orbit effect in band splitting. Similar is observed in CB- $K$ point, with almost twice splitting in 9R than 3R.  In the case of 9R stacking, which differs from 3R and 2H, we observed band splitting at VBM $\Gamma$ in the range of 46 meV to 109 meV across various TMDS but not observed in both 2H and 3R. This splitting at VBM $\Gamma$ in 9R is solely due to layer coupling as spin-orbit coupling has no contribution to it as shown in Fig. \ref{fig:9r_bandstructure_spin_orbit} and as detailed in Table \ref{tab:band_splitting}. The bandstructure of MoS$_2$ with opposite spin channels at $K$ and $K'$ for 3R and 9R is shown in Supplementary Information Fig. \ref{fig:3r_bandstructure_spin_channels} and \ref{fig:9r_bandstructure_spin_channels}.

\begin{table}[htbp]
\centering
\begin{tabular}{|c|r|r|r|r|r|r|r|r|r|r|r|r|}

\hline
  \multicolumn{1}{|c|}{\multirow{2}[0]{*}{\textbf{{{TMDs} }}}} 
 & \multicolumn{3}{c|}{\textbf{CBM}} & \multicolumn{3}{c|}{\textbf{VB-K}} & \multicolumn{3}{c|}{\textbf{VB-$\Gamma$}} & \multicolumn{3}{c|}{\textbf{CB-K}}\\
 
  \cline{2-13}  
  
\multicolumn{1}{|c|}{} & \multicolumn{1}{c|}{{2H}}  & \multicolumn{1}{c|}{{ 3R}} & \multicolumn{1}{c|}{{  9R}}  & \multicolumn{1}{c|}{{ 2H}} & \multicolumn{1}{c|}{{ 3R}} & \multicolumn{1}{c|}{{  9R}} & \multicolumn{1}{c|}{{ 2H}} & \multicolumn{1}{c|}{{ 3R}} & \multicolumn{1}{c|}{{  9R}} & \multicolumn{1}{c|}{{ 2H}} & \multicolumn{1}{c|}{{ 3R}} & \multicolumn{1}{c|}{{  9R}} \\
 \hline \hline

MoS$_2$ & 0  & 51 & 54   & 200 &150 &240 & 0 & 0 & 46 & 2 & 40 & 70  \\
\hline
MoSe$_2$ & 0  & 30 & 26   & 410 &190 &270 & 0 & 0 & 40 & 40 & 50 & 80 \\
\hline
MoTe$_2$ & 0  & 20 & 16   & 310 &220 & 337 & 0 & 0 & 46 & 15 & 40 & 100 \\
\hline
WS$_2$ & 0  & 195 & 190   & 480 & 430&560& 0 & 0 & 56 & 40 & 40 & 130 \\
\hline
WSe$_2$  & 0  & 160 & 167   & 660 &447 &580& 0 & 0 & 109 & 22 & 50 & 120  \\
\hline
WTe$_2$  & 0  & 165 & 171  & 560 & 490&650& 0 & 0 & 59 & 80 & 40 & 140 \\
\hline

\end{tabular}

    \caption{Band splitting due to spin-orbit coupling in meV for different TMDs.}
    \label{tab:band_splitting}
\end{table}

\subsection{Possible experimental synthesis}
In this section, we delve into the potential experimental synthesis of the 9R phase. There have been reports of successfully synthesizing the 3R phase by various means, such as restacking 2H monolayers, as discussed in Ref. \cite{3R_synthesis}. Using similar methods, it is conceivable to create the 9R phase by directly stacking monolayers in the 2H phase configuration.

Furthermore, our previous research on Ni-doped MoS$_2$  \cite{Karkee2020} yielded the 9R phase as an outcome. In this case, the doping of 3R MoS$_2$ at the octahedral intercalation site induced a phase transition from 3R to 9R. This transition resulted in the formation of trigonal pyramidal intercalation sites. It's worth noting that while the Mo/S-atop tetrahedral intercalation is energetically favored in 3R MoS$_2$ for Ni-doping, the trigonal pyramidal site represents a metastable state that offers the nearest minima for unstable octahedral intercalation in the 3R structure. Indeed, it appears that a precise approach to transition metal doping in the 3R phase has the potential to induce a phase transition to the 9R configuration. This method could serve as an alternative route for synthesizing the 9R phase, offering a controlled and potentially versatile means of achieving the desired crystal structure. It underscores the importance of doping strategies in tailoring the properties and phase transitions of two-dimensional materials like MoS$_2$.

\begin{figure}[H]
\includegraphics[scale=0.5]{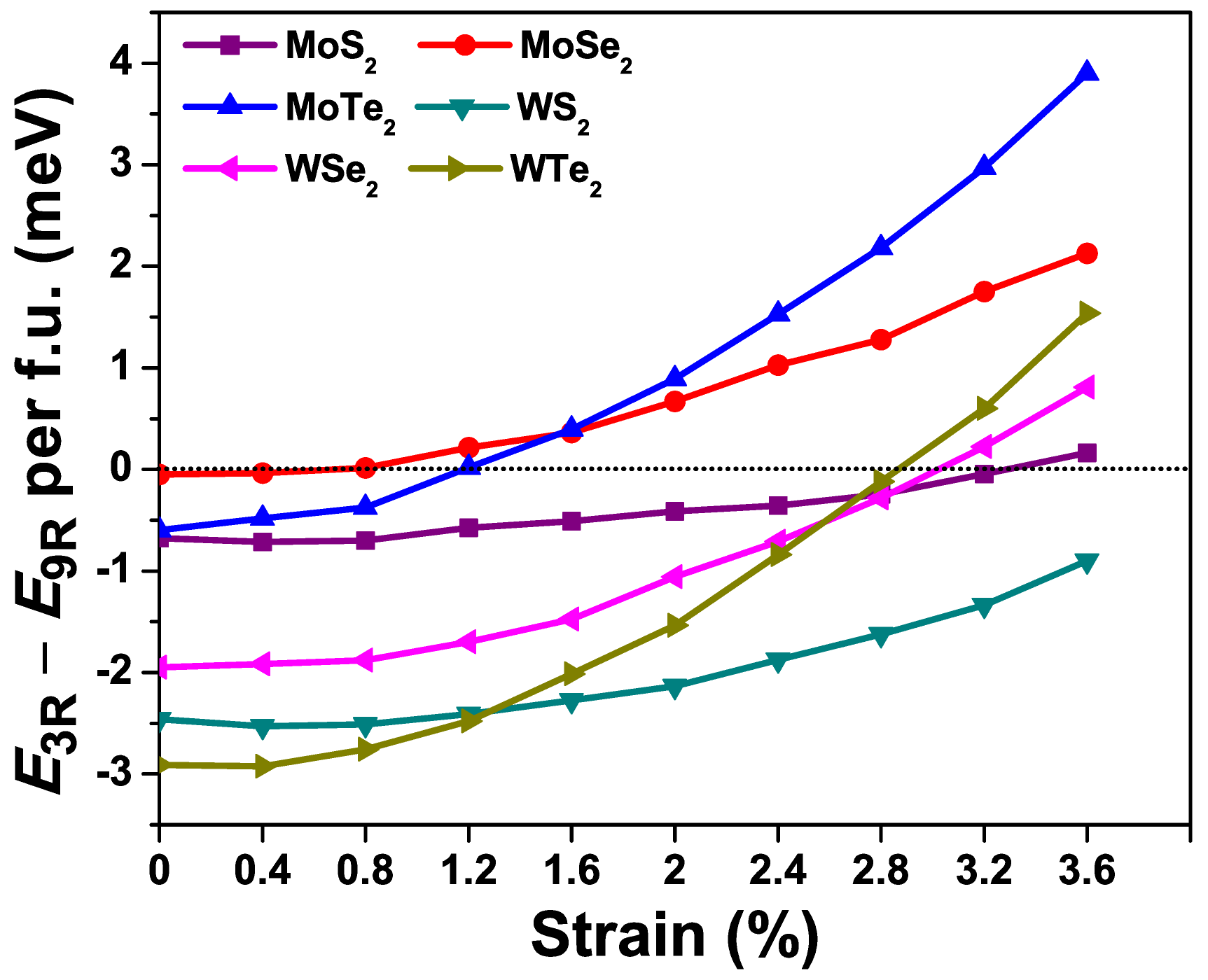}
\caption{Relative energy of 3R compared to 9R with (as described in Equation \ref{Strain_matrix}). Dotted horizontal line at 0  represents the strain at which 9R becomes more  energetically stable than 3R.}
\label{fig:3r_9r_energy_diff}
\end{figure}

Another approach that 9R may be synthesized is through shear strain on 3R. We found that under the strain $S$ defined by the following matrix:
\begin{equation}
  S =  
   \begin{pmatrix}
   1-\epsilon & 0 & 0\\ 
 0 & 1+\epsilon & -\frac{\epsilon}{2}\\ 
 0 &  -\frac{\epsilon}{2} & 1\\ 
 
   \end{pmatrix}
   \label{Strain_matrix}
\end{equation}
The total energy difference between 3R and 9R reduces to a small number or 9R becomes more energetically favorable, as shown in Figure \ref{fig:3r_9r_energy_diff}. This suggests, given the energy barrier between 3R and 9R being small, the perturbation in the structure due to external strain $S$ could favor 9R. We obtain matrix $S$ by minimizing the energy difference between two phases from equations \ref{strain_required1} and \ref{strain_required2}.
\begin{equation}
  U=\frac{V}{2}  \sum_{ij=1}^{6} \epsilon_{ij} C_{ij}
  \label{strain_required1}
\end{equation}

\begin{equation}
  U^{3R} - U^{9R}=\frac{1}{2}  \sum_{ij=1}^{6} \epsilon_{ij} (V^{3R} C_{ij}^{3R} - V^{9R} C_{ij}^{9R})
  \label{strain_required2}
\end{equation}

Here, $V$ is the volume per unit and $U$ is the energy. The difference $U^{3R} - U^{9R}$ is minimum when the eigenvalues of $V^{3R} C_{ij}^{3R} - V^{9R} C_{ij}^{9R}$ is minimum, and the corresponding eigenvectors defines the required strain for a given common factor $\epsilon$.

We also tested with non-shear strains such as compressive or tensile strains, hydrostatic  and  biaxial compression or elongation, but we found energy difference between 3R and 9R go up instead of down. We also tested the phase stability at higher temperature with harmonic approximation by calculating the lattice vibrations’ contributions to the free energy  $F = E - TS$, using the entropy as: \cite{lattice} 
\begin{equation}
    S(T)= k_B \sum_\lambda n_B(\hbar\omega_\lambda) \ln n_B (\hbar\omega_\lambda)
\end{equation}
where, $n_B$ is Bose-Einstein, $\omega$ is phonon frequency, $E$ is total energy (electronic and vibrational) and T is temperature. We found at higher temperature, energy difference increases further making 2H more stable than 3R and 9R, and 3R is more stable than 9R, as shown in example case of MoS$_2$ in Supplementary Information Fig. \ref{fig:Helmholz_free_energy}. 

\section{Conclusion}
In this research paper, we have introduced a novel nine-layer transition metal dichalcogenide (referred to as 9R) characterized by its dynamic and elastic stability. Notably, this phase exhibits Raman activity within the low-frequency range, a distinguishing feature not observed in the more established rhombohedral 3R phase. The 9R phase demonstrates superior piezoelectric properties, especially with regards to coefficients $d_{15}$ and $d_{22}$, and also displays a greater band splitting at the conduction band minimum compared to the 3R phase. We have also proposed several potential methods for synthesizing this phase, including direct stacking of monolayer 1H, inducing phase changes through transition metal doping, and applying shear strain. The introduction of the novel stacking sequence in 9R brings forth opportunities for exploring various applications related to stacking sequences. For instance, a study in Ref. \cite{D1NR03284D} has demonstrated the significance of stacking sequences in improving the hydrogen evolution reaction. Additionally, applications that arise from the breaking of inversion symmetry in this context demands further exploration.

\section*{Acknowledgments}
This work was supported by the National Science Foundation under Grant No. DMR-2144317. This work used computational resources from Pinnacles and Multi-Environment Computer for Exploration and Discovery (MERCED) clusters at UC Merced, funded by National Science Foundation Grants No. OAC-2019144 and ACI-1429783, and the National Energy Research Scientific Computing Center (NERSC), a U.S. Department of Energy Office of Science User Facility operated under Contract No. DE-AC02-05CH11231.

\section{Supplementary Information}

	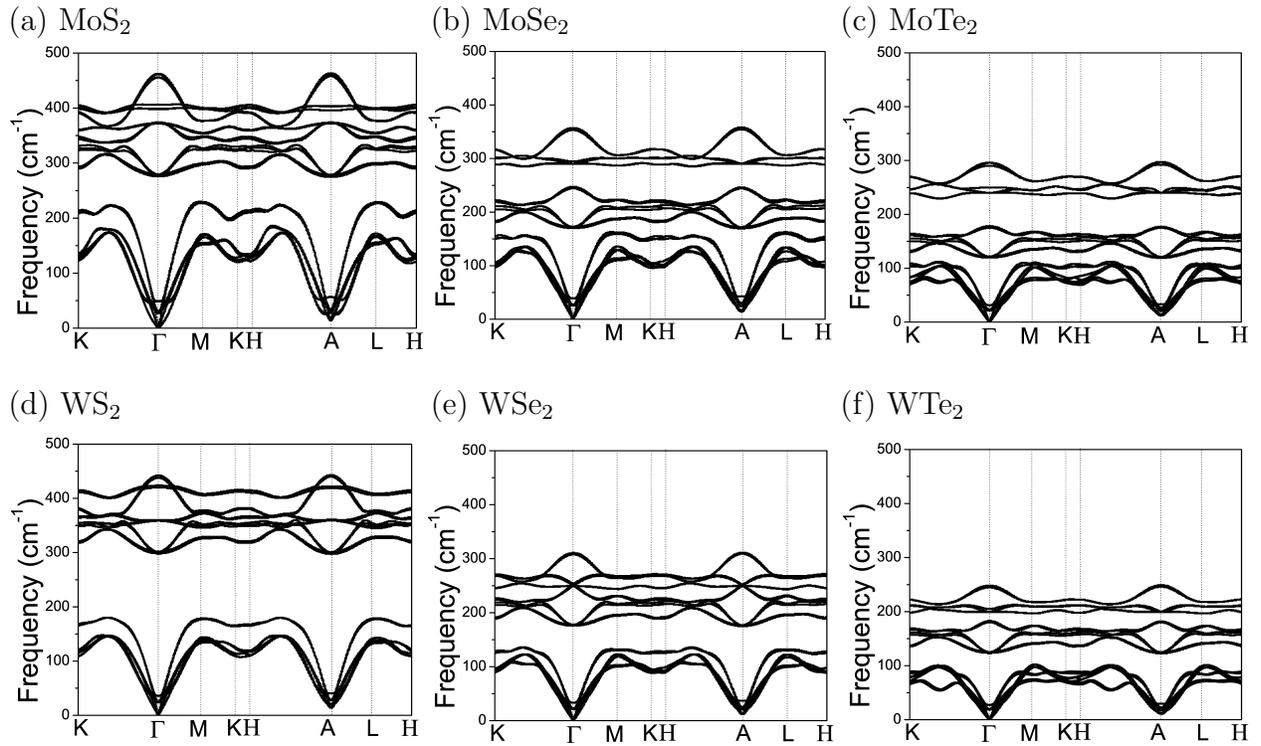
\begin{figure}[H]
			\centering{
			\begin{tikzpicture}
			\node [anchor=north west] (imgA) at (-0.15\linewidth,.58\linewidth){\includegraphics[width=0.335\linewidth]{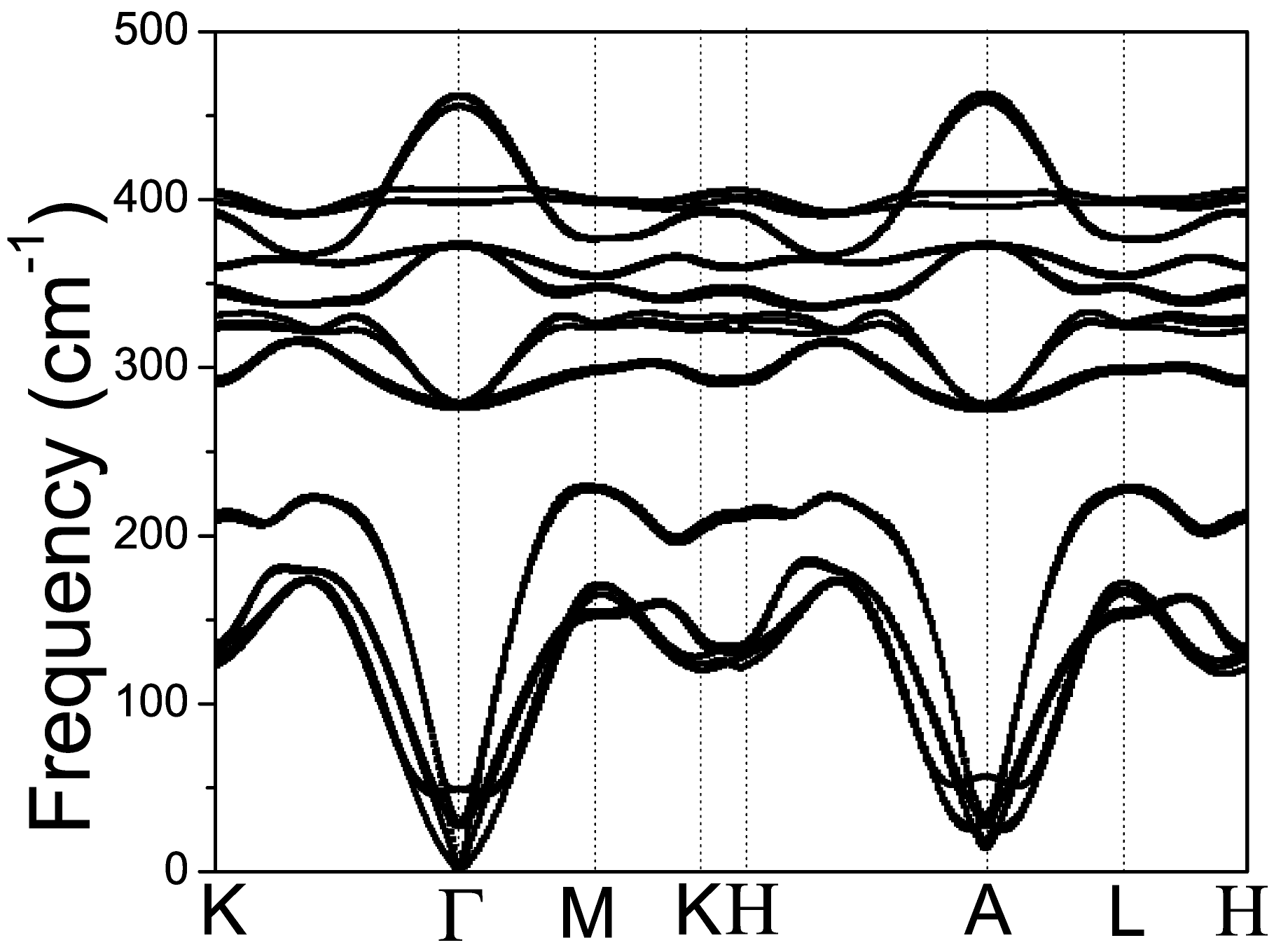}};
            \node [anchor=north west] (imgB) at (0.186\linewidth,.58\linewidth){\includegraphics[width=0.33\linewidth]{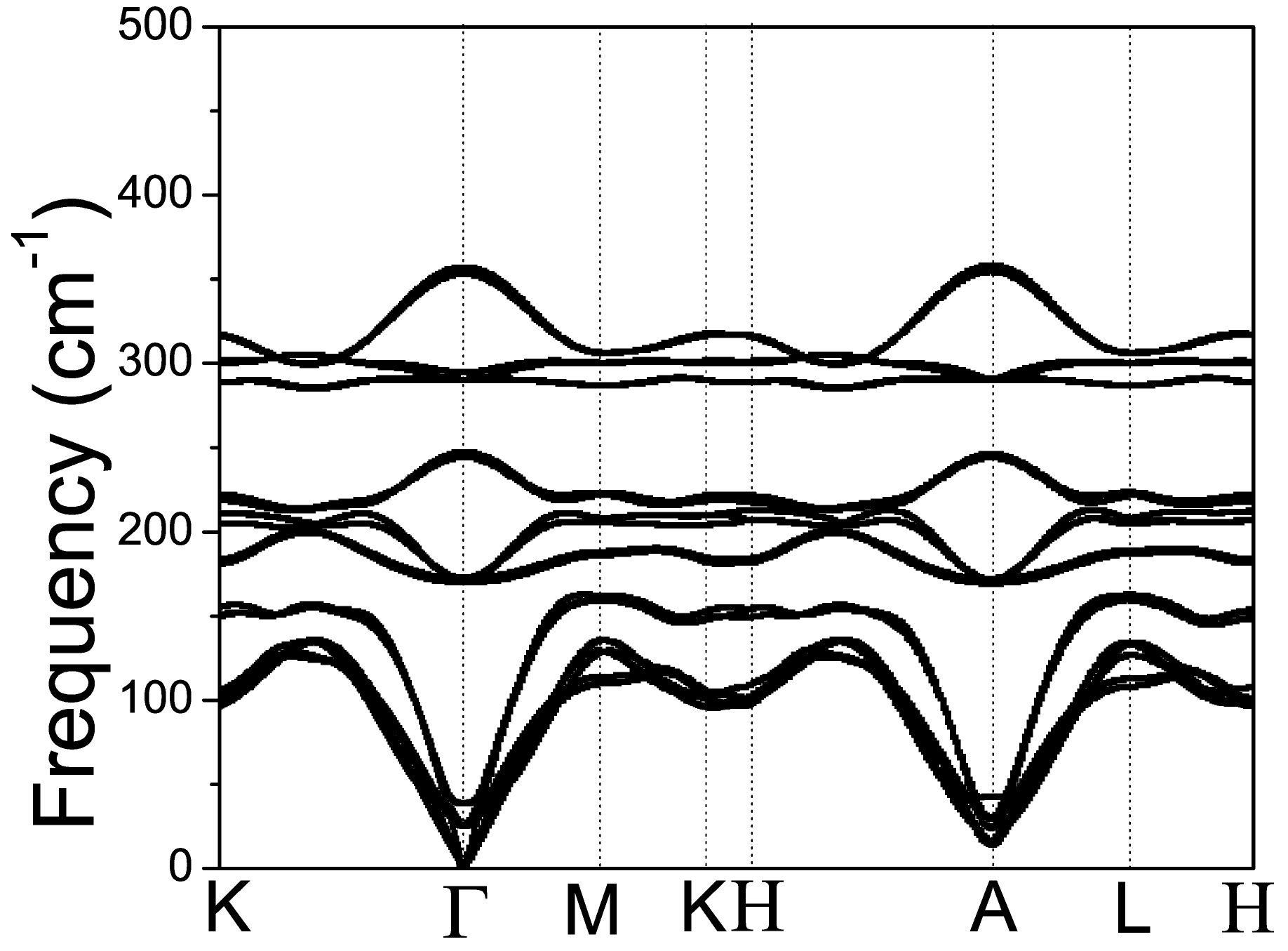}};
          \node [anchor=north west] (imgC) at (0.52\linewidth,.58\linewidth){\includegraphics[width=0.33\linewidth]{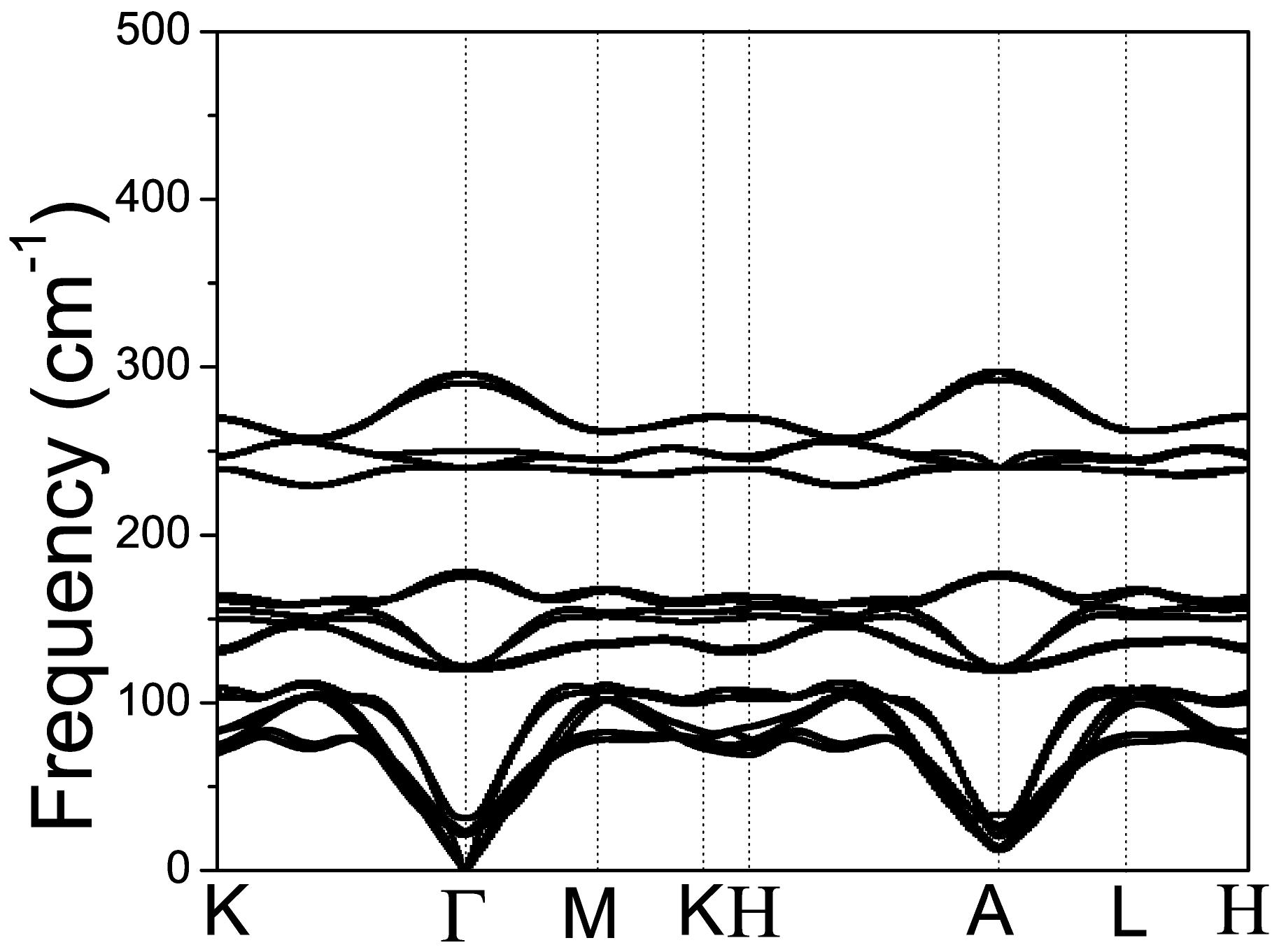}};
          
          	\node [anchor=north west] (imgD) at (-0.15\linewidth,.265\linewidth){\includegraphics[width=0.333\linewidth]{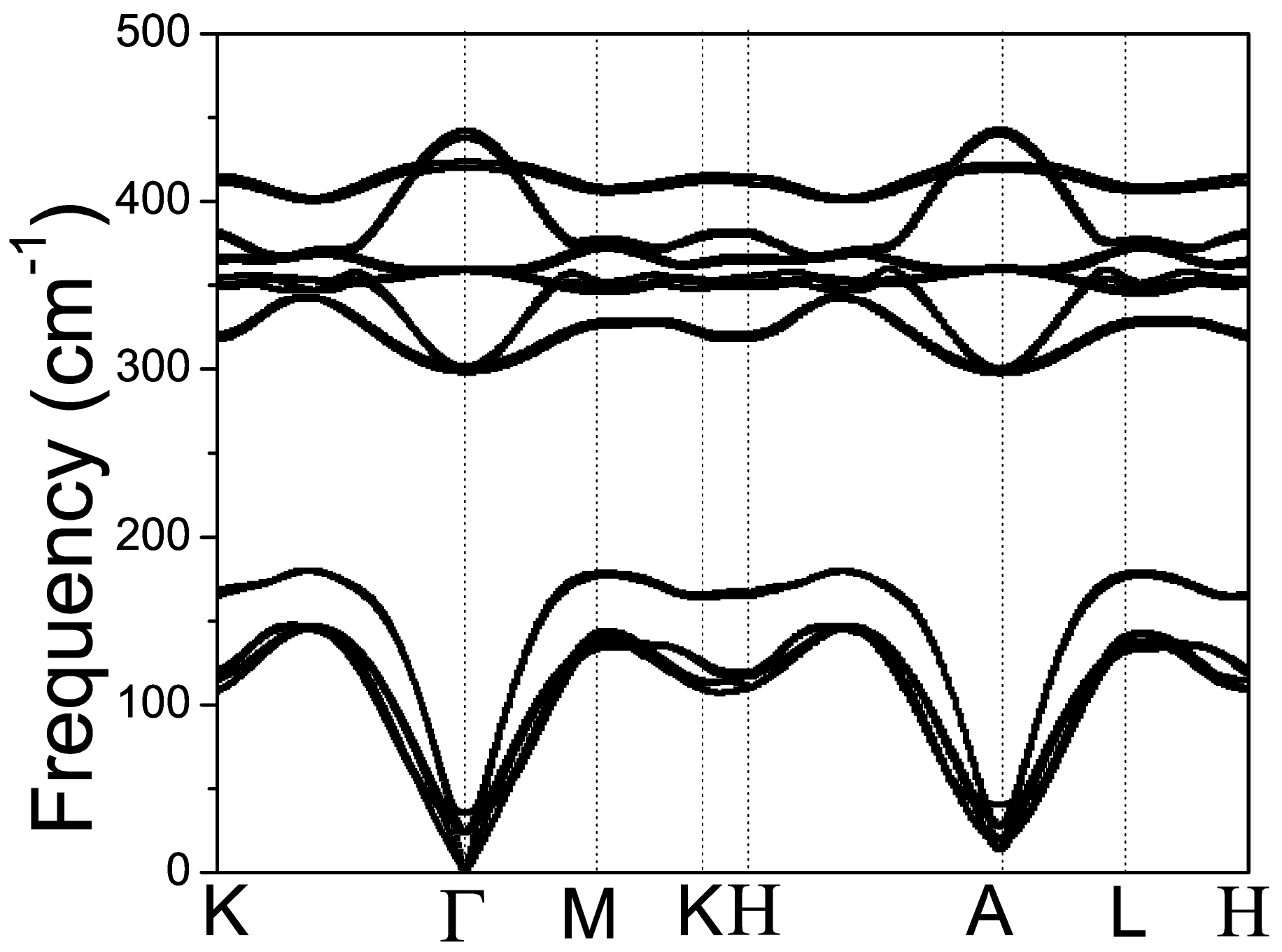}};
            \node [anchor=north west] (imgE) at (0.186\linewidth,.26\linewidth){\includegraphics[width=0.33\linewidth]{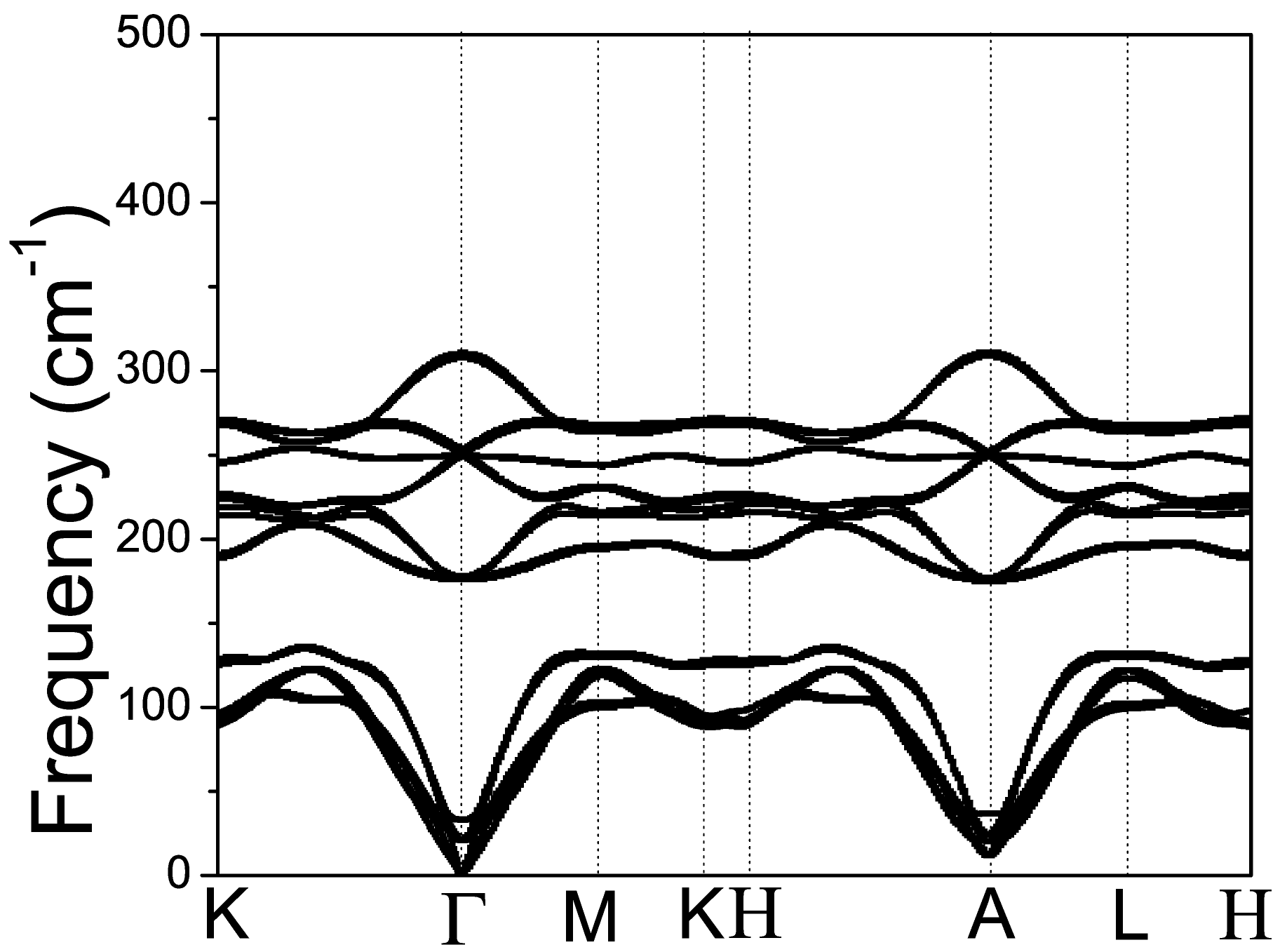}};
          \node [anchor=north west] (imgF) at (0.52\linewidth,.26\linewidth){\includegraphics[width=0.33\linewidth]{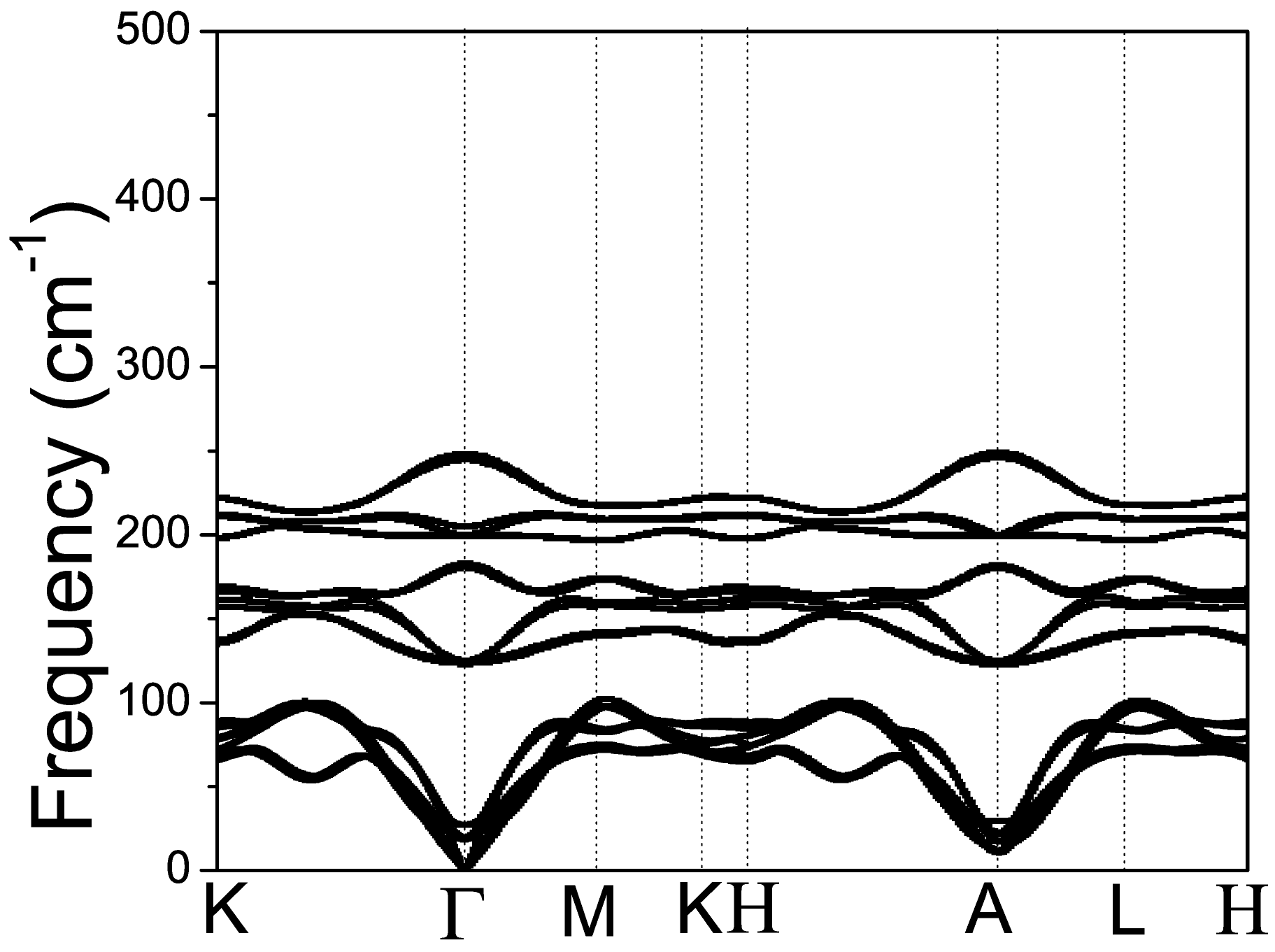}};
          
            \draw [anchor=north west] (-0.15\linewidth, .61\linewidth) node {(a) {\fontfamily{Arial}\selectfont \textbf{MoS$_2$}}};
            \draw [anchor=north west] (0.19\linewidth, .61\linewidth) node {(b) {\fontfamily{Arial}\selectfont \textbf{MoSe$_2$}}};
           \draw [anchor=north west] (0.52\linewidth, .61\linewidth) node {(c) {\fontfamily{Arial}\selectfont \textbf{MoTe$_2$}}};
            \draw [anchor=north west] (-0.15\linewidth, .30\linewidth) node {(d) {\fontfamily{Arial}\selectfont \textbf{WS$_2$}}};
            \draw [anchor=north west] (0.19\linewidth, .30\linewidth) node {(e) {\fontfamily{Arial}\selectfont \textbf{WSe$_2$}}};
           \draw [anchor=north west] (0.52\linewidth, .30\linewidth) node {(f) {\fontfamily{Arial}\selectfont \textbf{WTe$_2$}}};          
            \end{tikzpicture}
			}
			
			\caption{Calculated phonon bandstructure of 3R TMDs both in the in-plane and out-plane of the Brillouin zone.}
		\label{fig:3R_phonon}
			
		\end{figure}

	\begin{figure}[H]
			
			\centering{
			\begin{tikzpicture}
			\node [anchor=north west] (imgA) at (-0.15\linewidth,.58\linewidth){\includegraphics[width=0.33\linewidth]{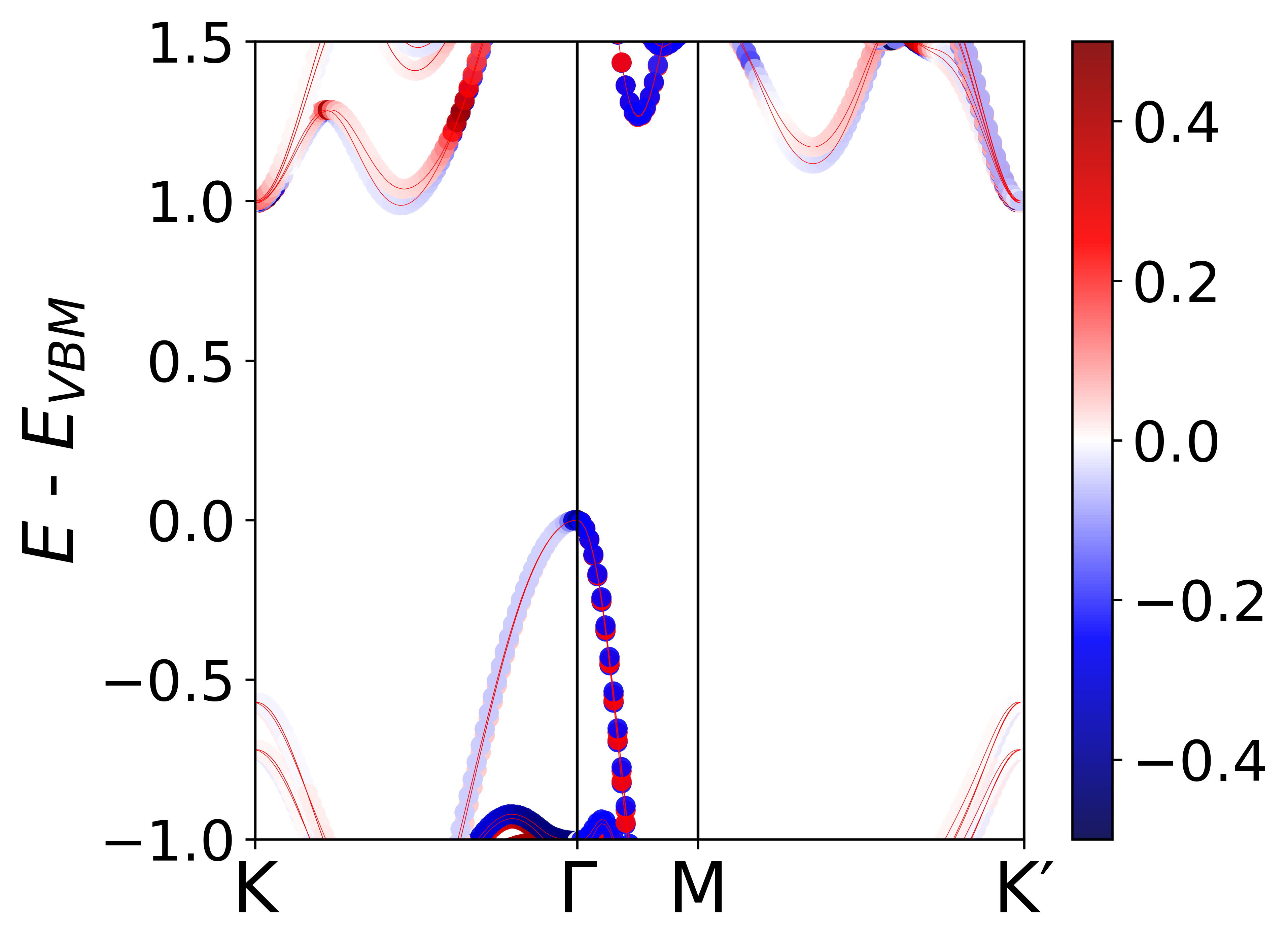}};
            \node [anchor=north west] (imgB) at (0.186\linewidth,.58\linewidth){\includegraphics[width=0.33\linewidth]{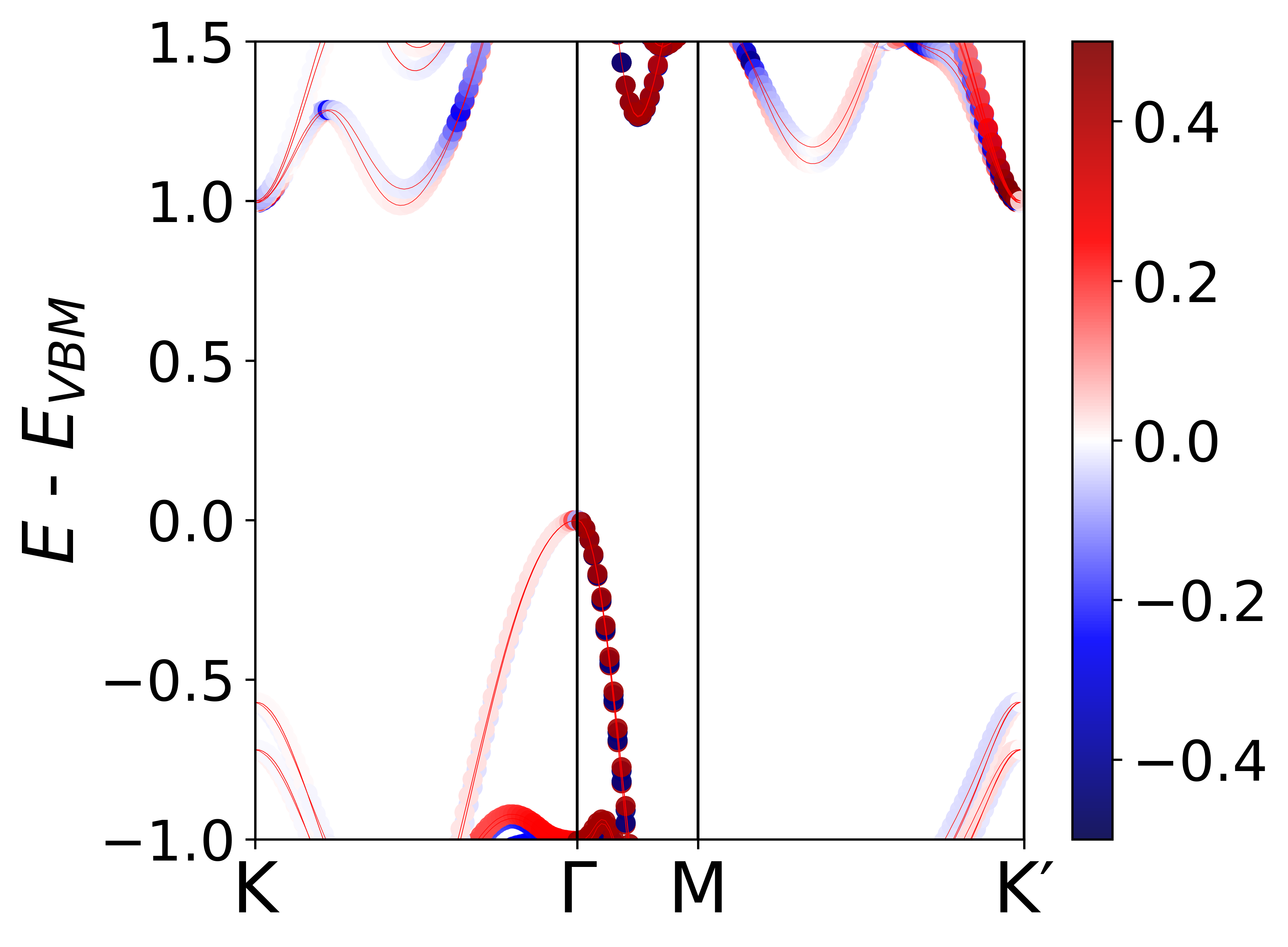}};
          \node [anchor=north west] (imgC) at (0.52\linewidth,.5895\linewidth){\includegraphics[width=0.33\linewidth]{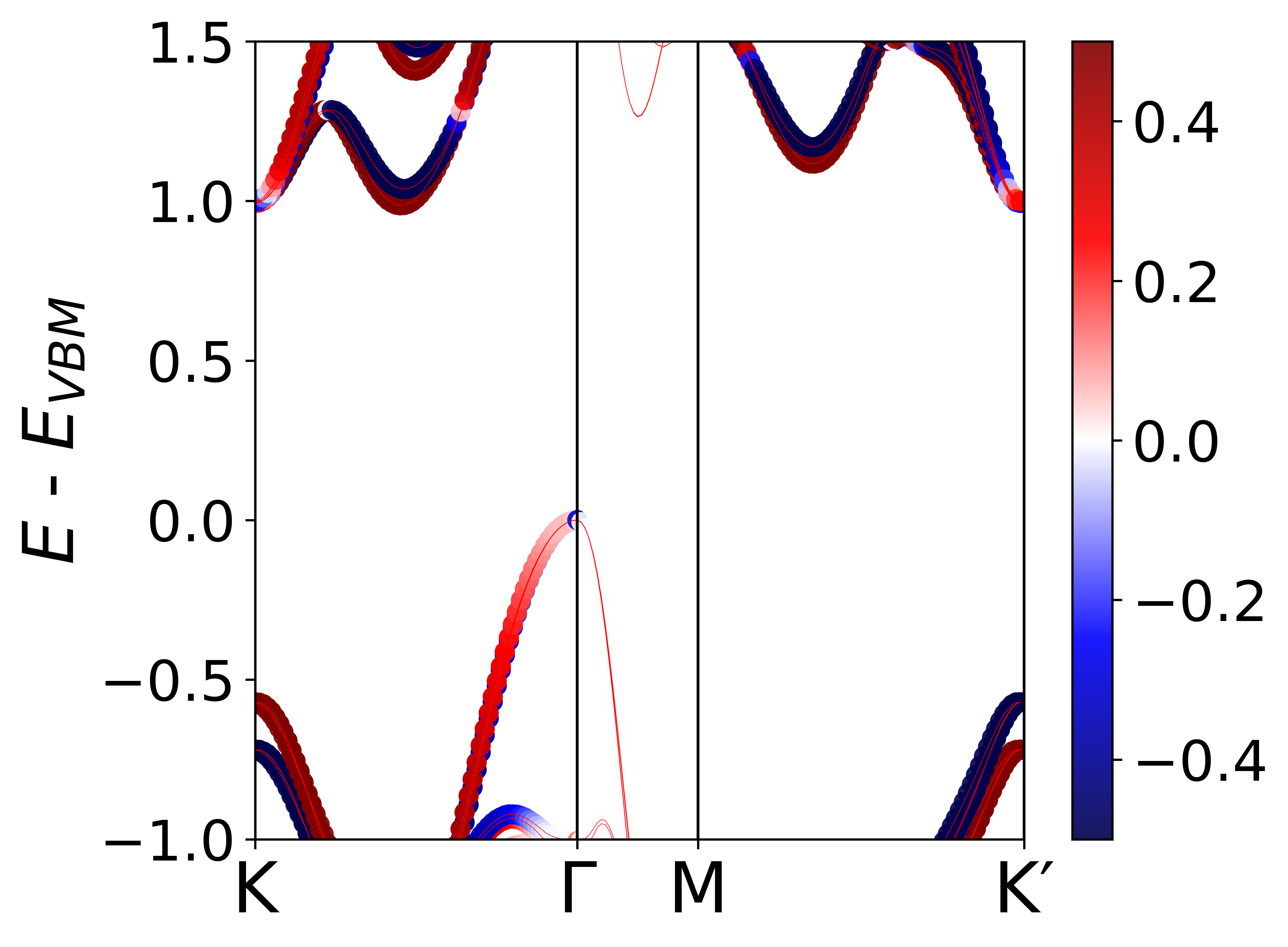}};

            \draw [anchor=north west] (-0.15\linewidth, .61\linewidth) node {(a) {\fontfamily{Arial}\selectfont \textbf{$<S_x>$}}};
            \draw [anchor=north west] (0.19\linewidth, .61\linewidth) node {(b) {\fontfamily{Arial}\selectfont \textbf{$<S_y>$}}};
           \draw [anchor=north west] (0.52\linewidth, .61\linewidth) node {(c) {\fontfamily{Arial}\selectfont \textbf{$<S_z>$}}};
      
            \end{tikzpicture}
			}
			
			\caption{Electronic bandstructure of 3R MoS$_2$ showing spin projections.}
		\label{fig:3r_bandstructure_spin_channels}
			
		\end{figure}

\begin{figure}[H]
    \centering
    \includegraphics[scale=0.6]{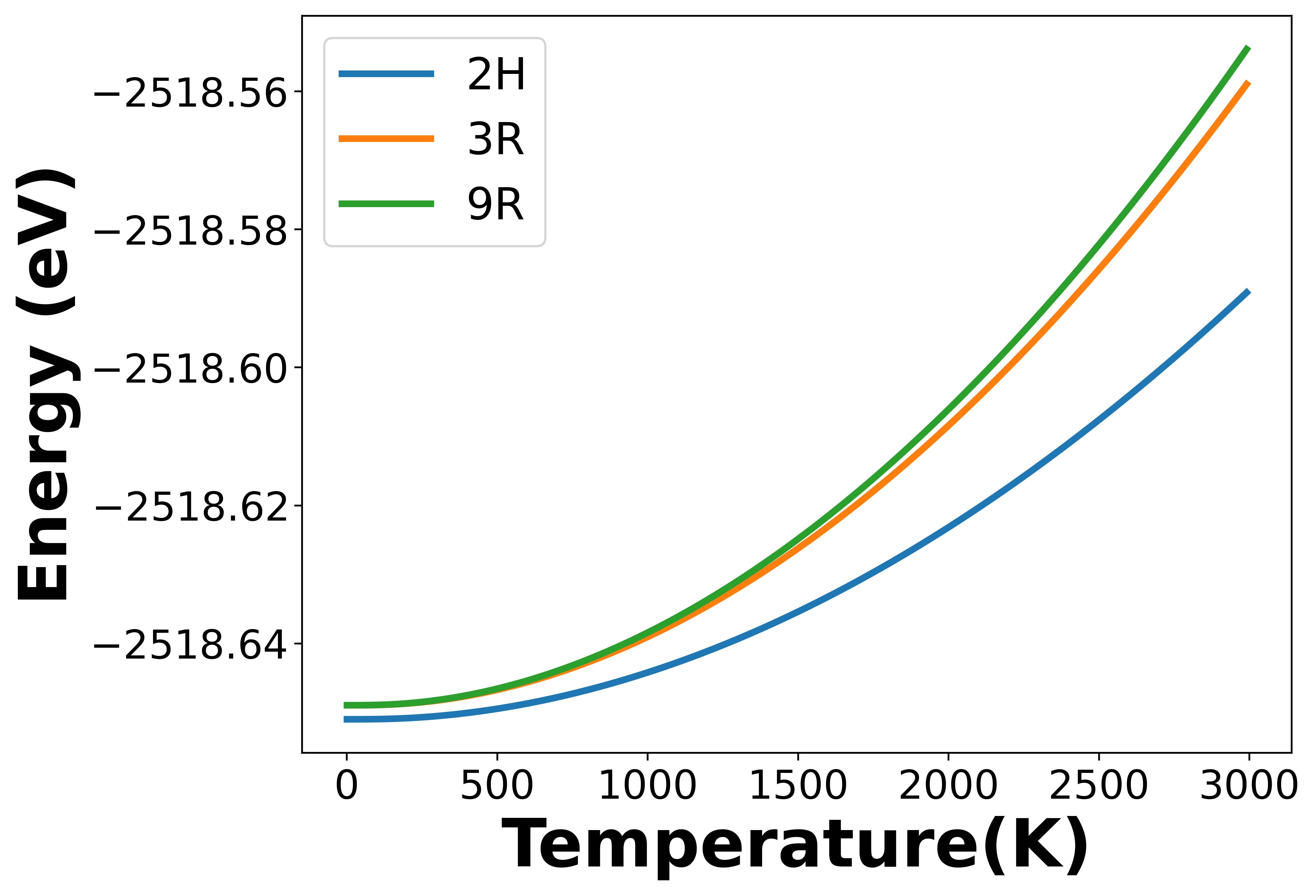}
    \caption{Free energy calculation of 2H, 3R and 9R MoS$_2$ as a function of temperature.}
    \label{fig:Helmholz_free_energy}
\end{figure}


\begin{table}[H]

\centering
\caption{Piezoelectric coefficients. The $d$ coefficients are in units of pm/V and $e$ coefficients are in units of C/m$^2$.}
\label{Tab:piezo tensor}
\begin{tabular}{|cc|r|r|r|r|r|r|r|r|}
\hline
\multicolumn{2}{|c|}{Phase} &$e_{15}$ & $d_{15}$ & $e_{22}$ & $d_{22}$ & $e_{31}$ & $d_{31}$ & $e_{33}$ & $d_{33}$  \\
\hline
\hline
 MoS$_2$ & 3R & -0.035 & -0.909 & 0.771 & 4.201 & -0.021 & 0.545 &-0.172 &  -3.650 \\
  & 9R &-0.182 & -9.708 &0.840 & 4.540 &-0.008 & 0.194 & -0.060 & -1.431  \\
\hline
\hline

 MoSe$_2$ & 3R & -0.111 & -5.510 &0.713 & 4.660 &-0.058 & 1.410 & -0.328 & -8.370  \\
  & 9R &-0.242 & -14.26 &0.816 & 5.380 &-0.021 & 0.465 &-0.112 & -2.830  \\
\hline
\hline
 
 MoTe$_2$ & 3R &-0.470 &-23.92 & 0.774 & 6.701 &-0.026 & 1.807 & -0.210 & -6.627  \\
  & 9R &-0.476 & -28.41 & 0.815 & 5.252 & -0.005 & 0.464 & -0.070 & -1.910  \\
\hline
\hline
 
 WS$_2$ & 3R &-0.101 & -5.280 & 0.560 & 2.836 & -0.017 & 0.425 & -0.135 & -3.324  \\
  & 9R & -0.071 & -4.197 & 0.523 & 2.528 & -0.006 & 0.150 & -0.047 & -1.162  \\
\hline
\hline

 WSe$_2$ & 3R & -0.040 & -3.384 & 0.514 & 3.137 & -0.051 & 1.346 & -0.306 & -7.945  \\
  & 9R & -0.131 & -8.575 & 0.523 & 3.060 & -0.004 & 0.213 & -0.041 & -1.060  \\
\hline
\hline
 
 WTe$_2$ & 3R & -0.173 & -9.376 & 0.601 & 4.824 & -0.009 & 1.928 & -0.171 & -5.587  \\
  & 9R & -0.257 & -15.33 & 0.810 & 6.747 & -0.004 & -0.512 & -0.050 & -1.607  \\
\hline

\end{tabular}
\end{table}

\bibliography{References}
\end{document}